\documentclass[aps,twocolumn,superscriptaddress,showpacs]{revtex4-1}
\usepackage{amsmath,amssymb,multirow}
\usepackage{graphicx}
\usepackage{MnSymbol}
\usepackage[usenames]{color}
\begin{document}
\title{Fate of the spin-$\frac{1}{2}$ Kondo effect in the presence of temperature gradients}
\author{Miguel A. Sierra}
\affiliation{Instituto de F\'{\i}sica Interdisciplinar y Sistemas Complejos
IFISC (UIB-CSIC), E-07122 Palma de Mallorca, Spain}
\author{Rosa L\'opez}
\affiliation{Instituto de F\'{\i}sica Interdisciplinar y Sistemas Complejos
IFISC (UIB-CSIC), E-07122 Palma de Mallorca, Spain}
\author{David S\'anchez}
\affiliation{Instituto de F\'{\i}sica Interdisciplinar y Sistemas Complejos
IFISC (UIB-CSIC), E-07122 Palma de Mallorca, Spain}
\begin{abstract}
We consider a strongly interacting quantum dot connected to two leads held at quite different temperatures. Our aim is to study the behavior of the Kondo effect in the presence of large thermal biases. We use three different approaches, namely, a perturbation formalism based on the Kondo Hamiltonian, a slave-boson mean-field theory for the Anderson model at large charging energies and a truncated equation-of-motion approach beyond the Hartree-Fock approximation. The two former formalisms yield a suppression of the Kondo peak for thermal gradients above the Kondo temperature, showing a remarkably good agreement despite their different ranges of validity. The third technique allows us to analyze the full density of states within a wide range of energies. Additionally, we have investigated the quantum transport properties (electric current and thermocurrent) beyond linear response. In the voltage-driven case, we reproduce the split differential conductance due to the presence of different electrochemical potentials. In the temperature-driven case, we observe a strongly nonlinear thermocurrent as a function of the applied thermal gradient. Depending on the parameters, we can find nontrivial zeros in the electric current for finite values of the temperature bias. Importantly, these thermocurrent zeros yield direct access to the system's characteristic energy scales (Kondo temperature and charging energy).
\end{abstract}
\maketitle

\section{Introduction}

The Kondo effect is a paradigmatic example of strong many-body physics in condensed matter~\cite{hewson}. Although this effect was firstly observed in metals with magnetic impurities, quantum dots (QDs) offer new possibilites for the manipulation of such many-body states owing to the tunability of the relevant parameters that control the physics of the problem. Since the discovery of the Kondo effect in semiconductor QD systems   \cite{dgo156,smc540,jsc182,ssa764,wgv210}, interest has been spurred both experimentally and theoretically during the last decades. A prototypical setup showing Kondo physics consists of a QD (an artificial quantum impurity) connected to two reservoirs at temperature $T$ lower than a characteristic temperature scale dubbed Kondo temperature $T_{K0}$. At such low temperature, the electric transport becomes highly correlated and the QD magnetic moment becomes screened by the electrons in the leads. A many-body singlet between the conducting and the localized electrons forms due to antiferromagnetic correlations originated from higher order spin-flip tunneling processes. At equilibrium, these correlations generate a sharp resonance in the  local density of states around the Fermi energy $\varepsilon_F$ whose width is of the order of $k_BT_{K0}$ \cite{tkn178,lig452,cos251}. Importantly, the Kondo temperature depends highly on the system parameters, i.e., the position of the QD level relative to the Fermi energy, $\varepsilon_d -\varepsilon_F$, the charging energy of the electrons inside the dot $U$ and the tunnel amplitude coupling to the  reservoirs  $\mathcal{V}_{\alpha k}$ (where $\alpha$ labels the lead connected to the dot and $k$ the electronic wave number). When a voltage bias is applied between the left and right contacts, a current is driven through the QD. In this situation, the differential conductance shows a peak that mimics the Kondo resonance (zero bias anomaly or Abrikosov-Suhl resonance) \cite{silvano}. For applied voltages of the order of $k_BT_{K0}/e$, the peak splits and a strong dephasing destroys the Kondo resonance \cite{aro156}.

Temperature gradients can in turn induce an electrical current (i.e., a thermocurrent).
Although an increasing background temperature clearly results in a destruction of the Kondo many-body singlet \cite{hewson},
much less is known on the effect of a thermal bias. A straightforward consequence is the generation of thermoelectric effects,
which are essential for designing reliable devices able to convert waste heat into useful electric work~\cite{book}. Since miniaturization progress has led to better performances~\cite{ven01}, it is interesting to investigate thermopower in quantum dots.
In fact, the thermopower turns out to be more sensitive than the conductance as a function of the gate voltage (linear response)~\cite{molenkamp,dzurak}.
In the nonlinear regime of transport, Ref.~\cite{sta93} observed a sign change in the thermovoltage $V_{\rm th}$ with the applied thermal bias $\theta$. More recently, Svensson \emph{et al.}~\cite{sfa105}  found analogous nonlinear effects in $V_{\rm th}$ for nanowire QDs. The latter experiments where theoretically addressed by two of us \cite{mas115} in which the emergence of highly nonlinear effects in the thermocurrent $I(\theta)$ and thermovoltage $V_{\rm th}$ nicely agreed with the experimental results. A Hartree-Fock scheme using nonequilibrium Green's functions within the equation-of-motion formalism was employed to model the thermoelectric transport in the Coulomb blockade regime. The origin of nonlinear thermocurrents and thermovoltages is related to a change in the direction of the electrical flow  due to the different QD resonance contributions. As a consequence, the differential thermoelectric conductance plots present a butterfly shape. Later on, Svilans \emph{et al.}~\cite{art100} reported similar experimental findings. 

The purpose of this work is to go beyond the Coulomb blockade regime of previous works and to thoroughly investigate the nonlinear thermoelectric  properties of quantum dots in the Kondo regime. Our first aim is to determine the fate of the Kondo effect upon application of a thermal gradient. In particular, we will discuss how the position of the Kondo peak and its width get modified due to thermal biases. We address this problem by employing different theoretical approaches: nonequilibrium Green's function formalism (NEGF) with higher-order equation of motion (EOM), slave-boson mean-field theory (SBMFT) and perturbative approach. These schemes encompass the whole range of temperatures, well below the Kondo temperature and for temperatures of the order or higher than $T_{K0}$. We compare the different approximations and study their range of validity.
Importantly, we predict that the Kondo resonance is destroyed when the thermal bias $\theta$ is sufficiently strong.
This result is obtained using both perturbative and SBMFT approaches. Further, we are able to give a fully analytical expression for the Kondo quench within perturbation theory. Our second aim is to analyze the transport properties of the system driven out of the equilibrium with both voltage and thermal biases and to examine the current behavior for arbitrary values of $\theta$. Our main finding in this case is a strong nonlinear behavior in the EOM current characteristics due to the existence of the Kondo resonance and the single-particle peaks. This is a relevant result because it would allow to experimentally measure the system's energy scales from the nontrivial zeros of the current.

Despite the fact that nonlinear thermoelectrics~\cite{review} in Kondo-correlated systems is quite a new topic, there is already a number of interesting results in the last years. Most of them are focused on the study of the Seebeck coefficient $S=V_{\rm th}/\theta$ employing a large variety of methods: nonperturbative resonant tunneling approximation \cite{dbo576}, second-order perturbation theory for the onsite Coulomb interaction \cite{bin117}, slave-boson noncrossing approximation (NCA) \cite{mkr155}, generalized Keldysh-based NCA \cite{jaz075}, nonequilibrium Green's functions beyond Hartree-Fock \cite{nat244}, quantum~\cite{pra235} and auxiliary~\cite{dor16} master equation approaches and dual fermions renormalized perturbation theory \cite{skiRPT}. In the linear regime, experimental evidence of the influence of the Kondo effect in the thermopower of a QD was reported in Ref.~\cite{sch05}. Thermopower and strong correlations are the subject of Refs.~\cite{cos10,zho10}. The Seebeck coefficient of an Aharanov-Bohm interferometer with an embedded QD shows modulated sign and magnitude~\cite{kim03}. When coupled to ferromagnetic leads, the quantum dot exhibits spin-dependent thermopowers~\cite{kra06,wey13}.
Double quantum dots supporting molecular states are believed to display enhanced thermoelectric power~\cite{and11}. The spin Seebeck coefficient, which measures the spin current generated in response to a thermal gradient, shows interesting features when many-body interactions are taken into account~\cite{rej12,zit13,cor12}. It turns out that the thermopower is a reliable probe of correlations in QD systems with SU(4) Kondo symmetry~\cite{rou12}. Kondo physics is able to boost the power output and efficiency~\cite{jaz075,zim14}. The effect of spin-orbit coupling of the Rashba type is treated in Ref.~\cite{kar13}. Finally, we briefly mention exciting topics where thermoelectrics in Kondo artificial impurities play a significant role: relaxation dynamics~\cite{ken13}, orbital degrees of freedom~\cite{lim14,ye14}, universal ac thermopower~\cite{chi14}, hybrid devices connected to ferromagnetic and superconducting leads~\cite{woj14}, assisted hopping~\cite{too14} and different configurations such as parallel~\cite{don14} or side coupled double QDs~\cite{woj16}.

Our paper is organized as follows. In Sec.~\ref{Sec:Ham} we describe our model Hamiltonian. We then present in Sec.~\ref{Sec:Scaling} a perturbative analysis for the case where a thermal gradient is applied across a two-terminal quantum dot.  In Sec.~\ref{Sec:SBMFT} we employ the slave-boson mean-field theory and investigate how the width and position of the Kondo peak are modified under voltage and thermal bias. We compute the thermoelectric current when Kondo correlations are present. We treat the high and moderate temperature regimes by using the equation-of-motion approach in Sec.~\ref{Sec:NEGF}. We consider the limit of large Coulomb repulsion and analyze the nonlinear transport for large voltages and thermal biases.
Finally, we summarize our main findings in Sec.~\ref{Sec:Conclusions}.
 
\section{Model Hamiltonian}\label{Sec:Ham}

\begin{figure}[t!]
\centering
\includegraphics[width=0.49\textwidth,clip]{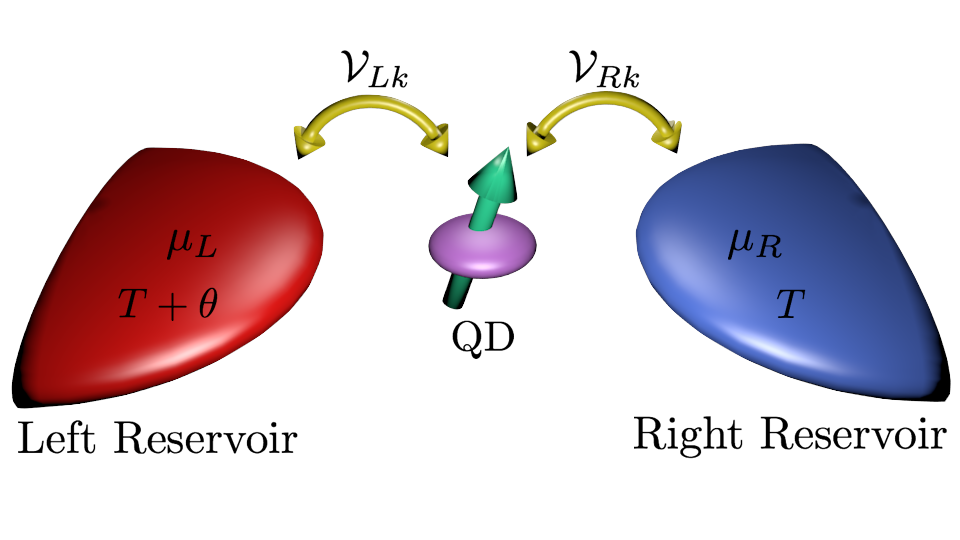}
\caption{(Color online) Sketch of the quantum dot system under the influence of a voltage ($\mu_L-\mu_R$)
and temperature gradient ($\theta$) applied between the 
two reservoirs. The system consists of two reservoirs (left $L$ and right $R$) connected through tunnel barriers (with tunneling amplitudes
$\mathcal{V}_{\alpha k}$) to an interacting quantum dot acting as an artificial quantum impurity (green arrow).}\label{fig:0}
\end{figure}

The setup under consideration is described by the single-impurity Anderson Hamiltonian \cite{ach420}. It describes the localized quantum dot level connected by tunneling barriers to two electronic reservoirs denoted with $\alpha=\{L,R\}$.  Each reservoir has an electrochemical potential $\mu_\alpha=\varepsilon_F+ eV_\alpha$ and a temperature $T_\alpha=T+\theta_\alpha $.
We hereafter consider that the left contact is held at a higher temperature than the right electrode
($\theta_L=\theta$, $\theta_R=0$ as in Fig.~\ref{fig:0}). Furthermore, we take
$\varepsilon_F=0$ as the reference energy and symmetric electric bias $V_L=-V_R=V/2$.

The total Hamiltonian is the sum of the following contributions:
\begin{eqnarray}\label{eq_H}
\mathcal{H}=\mathcal{H}_{\rm leads}+\mathcal{H}_{\rm dot}+\mathcal{H}_{\rm tun}\,.
\end{eqnarray}
The reservoir Hamiltonian reads 
\begin{equation}
\mathcal{H}_{\rm leads} = \sum_{\alpha k \sigma} \varepsilon_{\alpha k} C^\dagger_{\alpha k\sigma } C_{\alpha k \sigma}\,,
\end{equation}
where $\varepsilon_{\alpha k}$ is the energy dispersion relation for the $\alpha$ lead. Here, $C^\dagger_{\alpha k \sigma} $  ($C_{\alpha k \sigma} $) corresponds to the creation (annihilation) operator of an electron in the $\alpha$ lead with wave number $k$ and spin $\sigma=\{\uparrow,\downarrow\}$. The quantum dot system reads
\begin{equation}
 \mathcal{H}_{\rm dot}=\sum_\sigma \varepsilon_d d_{\sigma}^\dagger d_\sigma + U d_\uparrow^\dagger d_\uparrow d_\downarrow^\dagger d_\downarrow\,,
\end{equation} 
where the localized energy level is  $\varepsilon_d$ and the charging energy is $U$.  The tunnel Hamiltonian is represented as
\begin{equation}
\mathcal{H}_{\rm tun} = \sum_{\alpha k \sigma} \mathcal{V}_{\alpha k} C^\dagger_{\alpha k \sigma} d_{\sigma}+ {\rm H.c.}
\end{equation}
Here, $\mathcal{V}_{\alpha k}$ are the tunneling amplitudes. 

\section{Perturbative approach}\label{Sec:Scaling}

For the perturbative analysis, it is advantageous to resort to the Kondo 
Hamiltonian by means of a Schrieffer-Wolff transformation \cite{hewson}.  We remark that this Hamiltonian equivalence is only valid for the deep Kondo regime. Then, the Kondo Hamiltonian is $\mathcal{H}_{\rm K}=\mathcal{H}_{0}+\mathcal{H}_{1}$, where
\begin{equation}
\mathcal{H}_0=\sum_{\alpha k \sigma} \varepsilon_{\alpha k} C^\dagger_{\alpha k \sigma} C_{\alpha k \sigma}\,,
\end{equation}
and 
\begin{equation}\label{Eq:H1}
\mathcal{H}_1=\sum_{\alpha k \sigma \beta q s} \mathcal{J}_{\alpha \beta}(t)x_{\sigma s}C^\dagger_{\alpha k \sigma}C_{\beta q s} \,,
\end{equation}
where $\beta=\{L,R\}$ is a lead index and $x_{\sigma s}=\delta_{\sigma s}/4+\hat{S}_l s^l_{\sigma s}$
is defined in terms of  the quantum dot ($S_l$, $l=x,y,z$) and lead ($s_l$) spin operators, respectively.
The Kondo coupling,
\begin{eqnarray}
\mathcal{J}_{\alpha \beta}(t)=\mathcal{J}_{\alpha \beta}^{(0)}\exp{\left(-\frac{ie}{\hbar}[V_\alpha-V_\beta]t\right)} \,,
\end{eqnarray}
depends on the voltage difference between the two reservoirs
with $\mathcal{J}_{\alpha \beta}^{(0)}=-\mathcal{V}_{\alpha} \mathcal{V}_{\beta} U/[\varepsilon_d(U+\varepsilon_d)]$
(in this approach the $k$ dependence of the tunneling amplitudes is neglected).

We calculate the electrical conductance in perturbation theory up to third order in the Kondo coupling ($\mathcal{H}_1$) \cite{aka384,kam815}.  To do so, we first consider the expected value of the current operator,
\begin{eqnarray}\label{Eq:ISmatrix}
I=\langle \mathcal{S}(-\infty,0)\hat{I}(0)\mathcal{S}(-\infty,0)\rangle ,
\end{eqnarray}
with $\mathcal{S}(-\infty,0)$ the $S$ matrix given in terms of the perturbation $\mathcal{H}_1$,
\begin{equation}\label{S-matrix}
\mathcal{S}(-\infty,0) =\hat{T}\int_{-\infty}^{0} \mathcal{H}_{1}(\tau) d\tau\; .
\end{equation}
Here, $\hat T$ is the time-ordered operator. The explicit expression for the $\alpha$-current operator stems from the time derivative of the occupation number operator at lead $\alpha$, $\hat{I}_\alpha=-e\sum_{k\sigma}\partial_t ( C_{\alpha k\sigma}^\dagger C_{\alpha k\sigma} )$. After a few algebraic manipulations, we arrive at
\begin{eqnarray}\label{Eq:Ioperator}
\hat{I}(t)=\frac{ie}{\hbar}\sum_{kq\sigma s} \left(\mathcal{J}_{LR}(t)x_{\sigma s}C^\dagger_{Lk \sigma}C_{Rqs}-\text{H.c.}\right)\,.
\end{eqnarray}
The calculation of Eq.~(\ref{Eq:ISmatrix}) up to third order in the Kondo coupling by using Eq.~(\ref{S-matrix}) and Eq.~(\ref{Eq:Ioperator}) requires a lengthy algebra that is discussed in detail in Appendix \ref{Ap:ScalingPert}. After calculating the current expectation value, it is straightforward to obtain the expression for the electrical conductance $G\equiv dI/dV$
with $V=V_L-V_R$. We find
\begin{eqnarray}\label{conductance}
\nonumber G&=&-\frac{3e^2\pi}{4\hbar}\nu^2\left[\mathcal{J}_{LR}^{(0)}\right]^2\left(1-\frac{\nu}{2}\left(\mathcal{J}_{LL}^{(0)}+\mathcal{J}_{RR}^{(0)}\right)\ln{\left|\frac{k_B^2T_L T_R}{D_0^2}\right|}\right)\\
&&-\frac{e^2\pi}{4\hbar}\nu^2\left[\mathcal{J}_{LR}\right]^2 \; .
\end{eqnarray}
Here, $D_0=\sqrt{-\varepsilon_d (U+\varepsilon_d)}$ is the effective bandwidth,  and $\nu$ is the density of states of the leads, which we assume flat (wide band limit). The conductance $G$ has two contributions: the exchange cotunneling and the regular cotunneling terms [first and second lines in Eq.~(\ref{conductance}), respectively]. Notice that the logarithmic divergence contains the product of the reservoir temperatures. If we do not consider thermal gradients, our expression reduces to the conductance obtained in Ref.~\cite{kam815}.

The Kondo temperature is defined as the temperature at which the second order contribution in Eq.~(\ref{conductance}) dominates over the first term.
When there is neither thermal nor voltage biases applied to our system,
we recover the usual Kondo temperature \cite{fdm416} 
\begin{eqnarray}\label{Eq:ScalingTK}
k_B T_{K0}=D_0\exp{\left[\frac{\pi\varepsilon_d(U+\varepsilon_d)}{U\Gamma}\right]} \; .
\end{eqnarray}
Here $\Gamma=\Gamma_L+\Gamma_R$ is the total level broadening due to tunneling, with $\Gamma_\alpha=\pi \nu|\mathcal{V}_{\alpha}|^2$, which we take as a constant parameter.

In the presence of a temperature bias, we substitute $T_L=T+\theta$ and $T_R=T$ in Eq.~\eqref{conductance}. Therefore, the Kondo temperature becomes
 \begin{eqnarray}\label{tktheta}
T_K(\theta)=\sqrt{\left(\frac{\theta}{2}\right)^2+T_{K0}^2} -\frac{\theta}{2}\; .
\end{eqnarray}
This is a central result of our paper.
It yields the Kondo temperature as a function of the thermal gradient across a QD.
This $T_K$ should be understood as the energy scale at which the perturbative expansion fails in the presence of a thermal
gradient. It should not be confused with $T_{K0}$, which depends on QD intrinsic parameters.
\begin{figure}[b!]
\centering
\includegraphics[width=0.5\textwidth,clip]{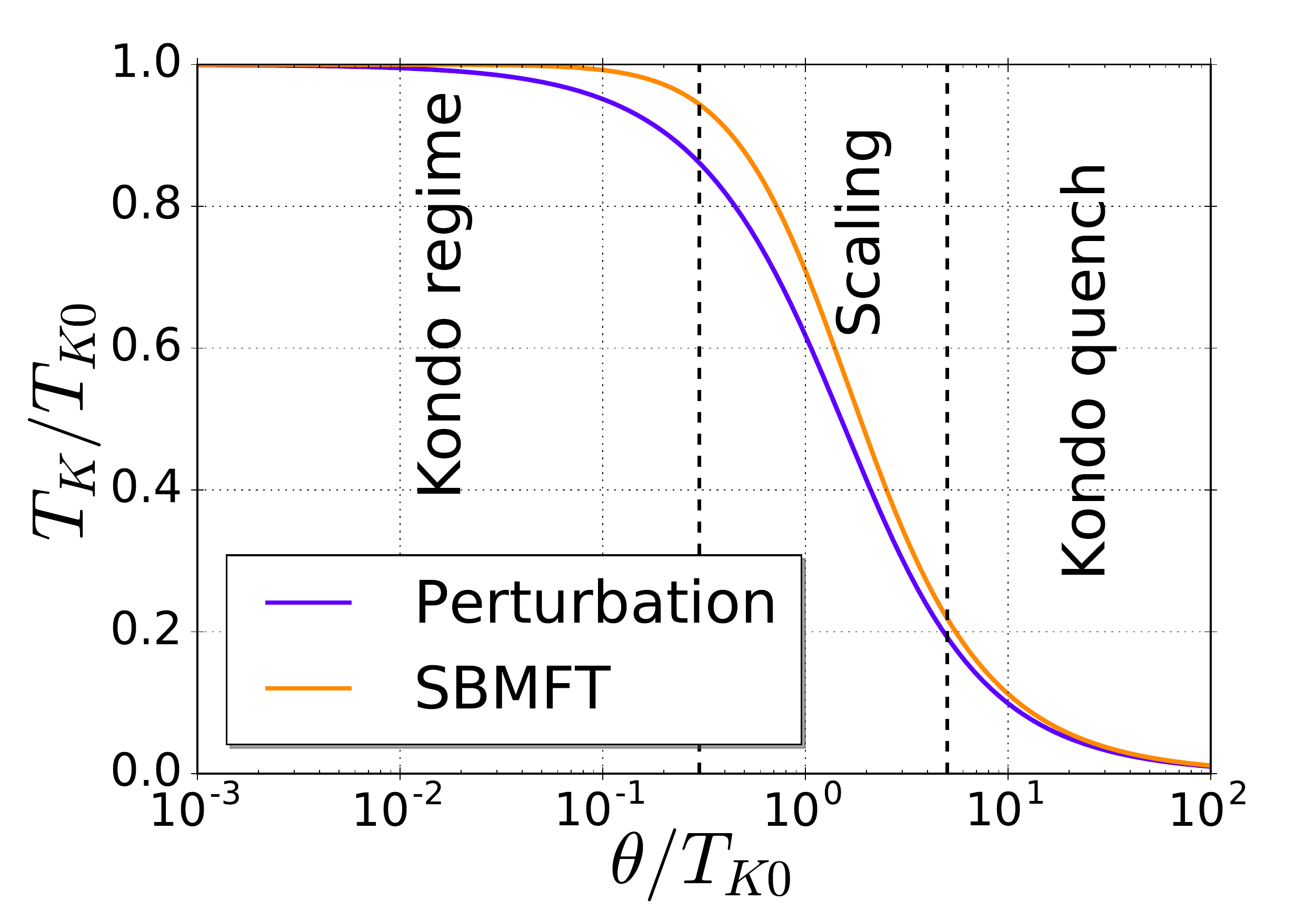}
\caption{(Color online). Normalized Kondo temperature $T_K/T_{K0}$ as a function of the thermal gradient $\theta/T_{K0}$ applied to a two-terminal quantum dot when $T_L=T+\theta$, and $T_R=T$. Blue line corresponds to the perturbative analysis result whereas orange line shows the Kondo temperature [$\tilde{\Gamma}=k_BT_K$ derived from Eq.~\eqref{Eq:25}] derived from slave boson mean-field theory. Here,  $T_{K0}=T_K(0)$ is defined as the intrinsic Kondo temperature.}\label{fig:13}
\end{figure}

Equation~(\ref{tktheta}) dictates that $T_K$ decreases as $\theta$ increases and eventually vanishes for very high values of the temperature bias. Our perturbative approach thus shows that a large thermal gradient kills the Kondo effect. However, Kondo correlations can survive for small values of $\theta$. These findings are illustrated in Fig.~\ref{fig:13} (see solid blue curve). Here, the normalized Kondo temperature $T_K/T_{K0}$ is displayed as a function of $\theta$.  We plot the Kondo temperature in a logarithmic scale for clarity. We observe that for small $\theta$ values the Kondo temperature stays roughly constant (Kondo regime) until $\theta$ becomes close to $T_{K0}$, where $T_K$ drops quickly (scaling region) and then vanishes monotonically (Kondo quench).

\section{Slave-boson mean-field theory} \label{Sec:SBMFT}

The perturbative approach presented above is only suitable for background temperatures higher than $T_{K0}$ because at low temperature the logarithmic dependence shown in Eq.~(\ref{conductance}) dominates and diverges. In order to extend our analysis toward the low temperature regime, we now analyze the same setup (a quantum impurity in the presence of large temperature biases) within slave-boson mean-field theory (SBMFT). Certainly, different formalisms are able to explain the Fermi liquid regime (e.g., renormalized perturbation theory \cite{ach007,aog988}). However, we choose to use SBMFT because it gives accurate results in the limit $T\to 0$.

Let us consider the Anderson Hamiltonian description [Eq.~(\ref{eq_H})] in the limit $U\rightarrow \infty$. This limit gives the correct low temperature behavior for the deep Kondo regime~\cite{pco303}. In the slave-boson formalism \cite{hewson} the dot operator $d_\sigma=b^\dagger f_\sigma$ is replaced by the product of a pseudofermion operator $f_\sigma$ and a boson field operator $b^\dagger $. When an electron with spin $\sigma$ is annihilated in the dot a vacuum state is created, which is represented by the boson creation operator $b^\dagger$ and the pseudofermion annihilation operator $f_\sigma$. Then, the tunneling Hamiltonian becomes
\begin{equation}
\mathcal{H}_{\rm tun}= \sum_{\alpha k \sigma} \mathcal{V}_{\alpha k} C^\dagger_{\alpha k \sigma} b^\dagger f_\sigma+ {\rm H. c.}
\end{equation}
To properly carry out the perturbation theory using $\mathcal{H}_{\rm tun}$ as a perturbation term, we need to rescale the tunneling amplitudes as  $\mathcal{V}_{\alpha k} \rightarrow \tilde{\mathcal{V}}_{\alpha k}\equiv\mathcal{V}_{\alpha k}/\sqrt{N}$ with $N$ representing the angular momentum degeneracy. The perturbation theory is done in the parameter $1/N$. Strictly when $N\rightarrow\infty$ the perturbation theory leads to exact results \cite{pco303}. We impose the condition $U\rightarrow\infty$ by adding a Lagrange multiplier $\lambda$ to the Hamiltonian as 
\begin{equation}\label{eqlag}
\mathcal{H}_{\rm Lag}=\lambda\left(b^\dagger b +\sum_\sigma f_\sigma^\dagger f_\sigma -1\right) \; .
\end{equation}
This term ensures that the Hilbert space does not contains the double occupancy dot state.

Next, we consider the mean-field solution of the Hamiltonian. This corresponds to the lowest order in the large-$N$ expansion. Then, the boson operator is replaced with its mean value $\langle b \rangle=\tilde{b}$, $\tilde{b}$ being a $c$-number. This way, charge fluctuations are completely screened out. This assumption is valid as long as $T<T_{K0}$ and the dot level lies within the Kondo regime (Fermi liquid fixed point). 
Our goal is then to derive the mean-field equations for both the expectation value of $b$ and the Lagrange multiplier. We first determine the equation of motion for $\tilde{b}$ in the stationary limit,
\begin{eqnarray}\label{SlvCond1}
\sum_{\alpha k \sigma} \tilde{\cal V}_{\alpha k} G^<_{f\sigma,\alpha k \sigma}(t,t)=-i N\lambda|\tilde{b}|^2/\hbar\,,
\end{eqnarray}
where $G^<_{f\sigma,\alpha k \sigma}(t,t')=-(i/\hbar) \langle C_{\alpha k \sigma }^\dagger(t') f_\sigma(t) \rangle $ is the lesser Green's function of the tunneling process~\cite{hewson2}. The second mean-field equation is directly the equation for the Lagrange multiplier that imposes the closure relation for the restricted Hilbert space [Eq.~\eqref{eqlag}] in terms of the mean-field value of the boson operator  
\begin{eqnarray}\label{SlvCond2}
\sum_\sigma G^<_{f\sigma,f\sigma}(t,t)=i(1-N|\tilde{b}|^2)/\hbar\,,
\end{eqnarray}
where $G^<_{f\sigma,f\sigma}(t,t')=-(i/\hbar) \langle f_{\sigma }^\dagger(t') f_\sigma(t) \rangle $
is the dot pseudofermion lesser Green's function.
The two nonlinear equations,  Eqs.~(\ref{SlvCond1}) and~(\ref{SlvCond2}), need to be solved self-consistently. By combining Eqs.~(\ref{SlvCond1}) and~(\ref{SlvCond2}) in a single complex equation and resorting to the Fourier space one gets
\begin{eqnarray}\label{SlvCondInt}
\frac{2}{\pi}\int_{-D}^Dd\omega \frac{\mathcal{F}(\omega)}{\omega-\tilde{\varepsilon}_d+i\tilde{\Gamma}}=(\varepsilon_d- \tilde{\varepsilon}_d)\frac{N}{\Gamma}-i\left(1-N\frac{\tilde{\Gamma}}{\Gamma}\right) \; ,
\end{eqnarray}
where $\tilde{\varepsilon}_{d}=\varepsilon_{d}+\lambda$ is the renormalized dot level position due to Kondo 
correlations,
$\tilde\Gamma= |\tilde{b}|^2 \Gamma$ is the tunnel hybridization also renormalized by Kondo correlations,
and $\mathcal{F}(\omega)$ is a nonequilibrium distribution function \cite{cab425} 
\begin{eqnarray}\label{effetivefermi}
\mathcal{F}(\omega)=\sum_{\alpha} \frac{\Gamma_\alpha f_\alpha(\omega)}{\Gamma}.
\end{eqnarray}
Here, $f_\alpha(\omega)=1/[1+\exp{\{(\omega-\mu_\alpha)/(k_BT_\alpha)\}}]$ is the Fermi-Dirac function of lead $\alpha$. Once the two mean-field parameters are determined, we are in a position to compute the QD spectral function  
\begin{eqnarray}
\rho_{d\sigma}(\omega)=-\frac{|\tilde{b}|^2}{\pi}\text{Im}{[G^r_{f\sigma,f\sigma}(\omega)]}\,,
\end{eqnarray}
 where $G^r_{f\sigma,f\sigma}(t,t')=-\frac{i}{\hbar}\theta(t-t')\langle [f^\dagger_\sigma(t') ,f_\sigma(t)]_+\rangle$ is the pseudofermion retarded Green's function in terms of $[\ldots]_+$, the anticommutator of two operators. Calculating the equation of motion for this Green's function, we find a closed system of equations that gives
\begin{eqnarray}\label{KondoPeakSlv}
\rho_{d\sigma}(\omega)=\frac{|\tilde{b}|^2}{\pi}\frac{\tilde{\Gamma}}{(\omega -\tilde{\varepsilon}_d)^2+\tilde{\Gamma}^2}\, .
\end{eqnarray}
Usually the Kondo resonance is pinned at the Fermi energy. Then, $\tilde{\varepsilon}_{d} \approx \varepsilon_F$, leading to the Kondo resonance.
The peak width is given by $\tilde{\Gamma}$.
We remark that Eq.~\eqref{KondoPeakSlv} obeys the Friedel sum rule, $\pi\Gamma \rho_{d\sigma}(\tilde{\varepsilon}_d)=1$, even though the energy integration is not able give the full occupation since SBMFT does not capture the single-particle peaks. SBMFT only describes the Kondo peak properly, which nevertheless suffices for our purpose of analyzing the Kondo temperature, as we next discuss.

\subsection{Kondo resonance}\label{sec:bifur}

Let us determine how small electrical and thermal biases modify the Kondo resonance within SBMFT. To obtain analytical results we consider  $ |\varepsilon_{d}|\gg\tilde{\varepsilon}_{d}$ (deep Kondo regime) together with $|\tilde{b}|^2\ll 1$ (Fermi liquid).
We denote the width of the Kondo resonance with $k_B T_K \equiv \tilde{\Gamma}$. Whereas the Kondo temperature $T_K$ found
in Eq.~\eqref{tktheta} is identified as the energy scale where perturbation theory breaks down in the presence of external biases,
here we can interpret $k_B T_K$ as the energy associated to the width of the Kondo peak, which also depends on the external biases.
Of course, at equilibrium and for $T=0$ we recover the intrinsic Kondo temperature given by Eq.~\eqref{Eq:ScalingTK} for $U\to\infty$:
\begin{eqnarray}\label{Eq:TK0SBMFT}
k_B T_{K0}\equiv\tilde{\Gamma}(\theta=0)=De^{-\pi|\varepsilon_d|/\Gamma}\; ,
\end{eqnarray} 
which depends on the QD parameters and the lead bandwidth $D$.

Our objective now is to examine the behavior of $k_B T_K \equiv \tilde{\Gamma}$ as a function of voltage and temperature shifts.
In the voltage-driven case, but still $T=0$, the nonequilibrium distribution function appearing in Eq.\ (\ref{SlvCondInt}) plays a fundamental role when $\tilde{\Gamma}$ is computed as a function of the voltage bias. In such case, the distribution function becomes a doubly stepped function depending on the voltage of the leads. Hence, one arrives at the following expressions:
\begin{eqnarray}
\tilde{\Gamma}(V)\tilde{\varepsilon}_d&=&0\; ,\\
\tilde{\varepsilon}_d^2-\left(\frac{eV}{2}\right)^2-[\tilde{\Gamma}(V)]^2&=&-(k_B T_{K0})^2\; .\label{SlvVsol}
\end{eqnarray}
Remarkably, when $eV$ approaches $2k_BT_{K0}$ the width of the Kondo peak drops to zero whereas $\tilde{\varepsilon}_d \approx 0$ due to the fact that $\tilde{b}\rightarrow 0$ and $\lambda=-\varepsilon_{d}$. Then, at exactly $eV=2k_BT_{K0}$ a phase transition occurs and the renormalized dot level position undergoes a bifurcation. Such a bifurcation indicates that the Kondo resonance splits due to the presence of the electrical bias. This result agrees with the calculations made in Refs.~\cite{pco205,lop04} and the experimental result shown in Ref.~\cite{silvano}. We note that SBMFT is valid for voltages smaller than $2k_BT_{K0}$ since the predicted phase transition is in reality a crossover. 

\begin{figure}[b!]
\centering
\includegraphics[width=0.5\textwidth,clip]{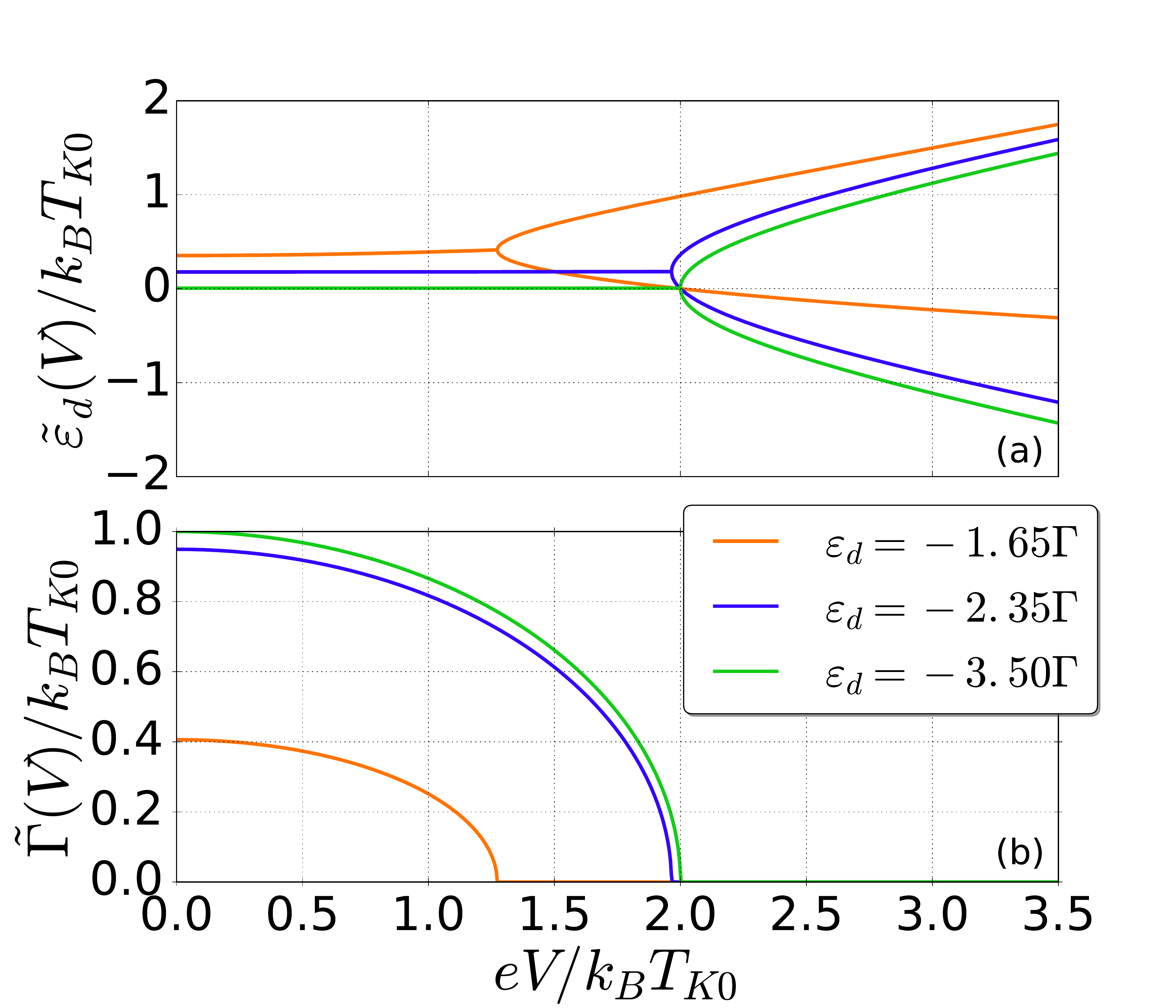}
\caption{Position of the SBMFT (a) renormalized energy level $\tilde{\varepsilon}_d$ and (b) width $\tilde{\Gamma}$ as a function of the applied voltage for different dot level positions. The case $\varepsilon_d=-3.5\Gamma$ agrees with the analytical result given by Eq.~\eqref{SlvVsol}. Parameters: $D=100\Gamma$, $k_B T = 0$ and $\Gamma_L=\Gamma_R=\Gamma/2$.}\label{fig:8}
\end{figure}

We present in Fig.~\ref{fig:8} our numerical results for Eq.~(\ref{SlvCondInt}) in the voltage-driven case and $T=0$. We observe that the numerical result nicely fits the analytical result found in Eq.~(\ref{SlvVsol}), especially for the deep Kondo regime (see the case for $\varepsilon_d=-3.5\Gamma$).  The phase transition occurs  when $eV=2k_BT_{K0}$, at which point the splitting of the Kondo resonance takes place [Fig.~\ref{fig:8}(a)]. When the dot level approaches towards the mixed-valence regime ($\varepsilon_d\simeq -\Gamma$), charge fluctuations become important and the phase transition occurs at much lower voltage bias values.
As expected, the Kondo resonance becomes narrower as voltage grows [Fig.~\ref{fig:8}(b)].
 
For the temperature-driven, case an analytical treatment can also be performed, leading to the result
\begin{equation}\label{Eq:25}
\sum_\alpha \frac{\Gamma_\alpha}{\Gamma} \left[\log{\left|\frac{2\pi k_B T_\alpha}{D}\right|}+\psi{\left(\frac{1}{2}+\frac{i\tilde{\varepsilon}_d+\tilde{\Gamma}}{2\pi k_B T_\alpha}\right)}\right]=\frac{\pi N \varepsilon_d}{2\Gamma}\; ,
\end{equation}
where $\psi(x)$ denotes the digamma function. By expanding the digamma function around $T=0$
we find the leading-order contribution $\tilde{\Gamma}$ within a temperature gradient expansion.
This amounts to considering a Sommerfeld expansion in the integral of Eq.~(\ref{SlvCondInt}).
However, we need to calculate the second order term to get a dependence of $\tilde{\Gamma}$ on the thermal gradient ,
\begin{eqnarray}
\tilde{\varepsilon}_d&=&0\; ,\\
\tilde{\Gamma}&=&k_BT_{K0} e^{-\frac{\pi^2}{12}\frac{T_L^2+T_R^2}{(T_{K0})^2}}\; .\label{TKvsT}
\end{eqnarray}

\begin{figure}[b!]
\centering
\includegraphics[width=0.5\textwidth,clip]{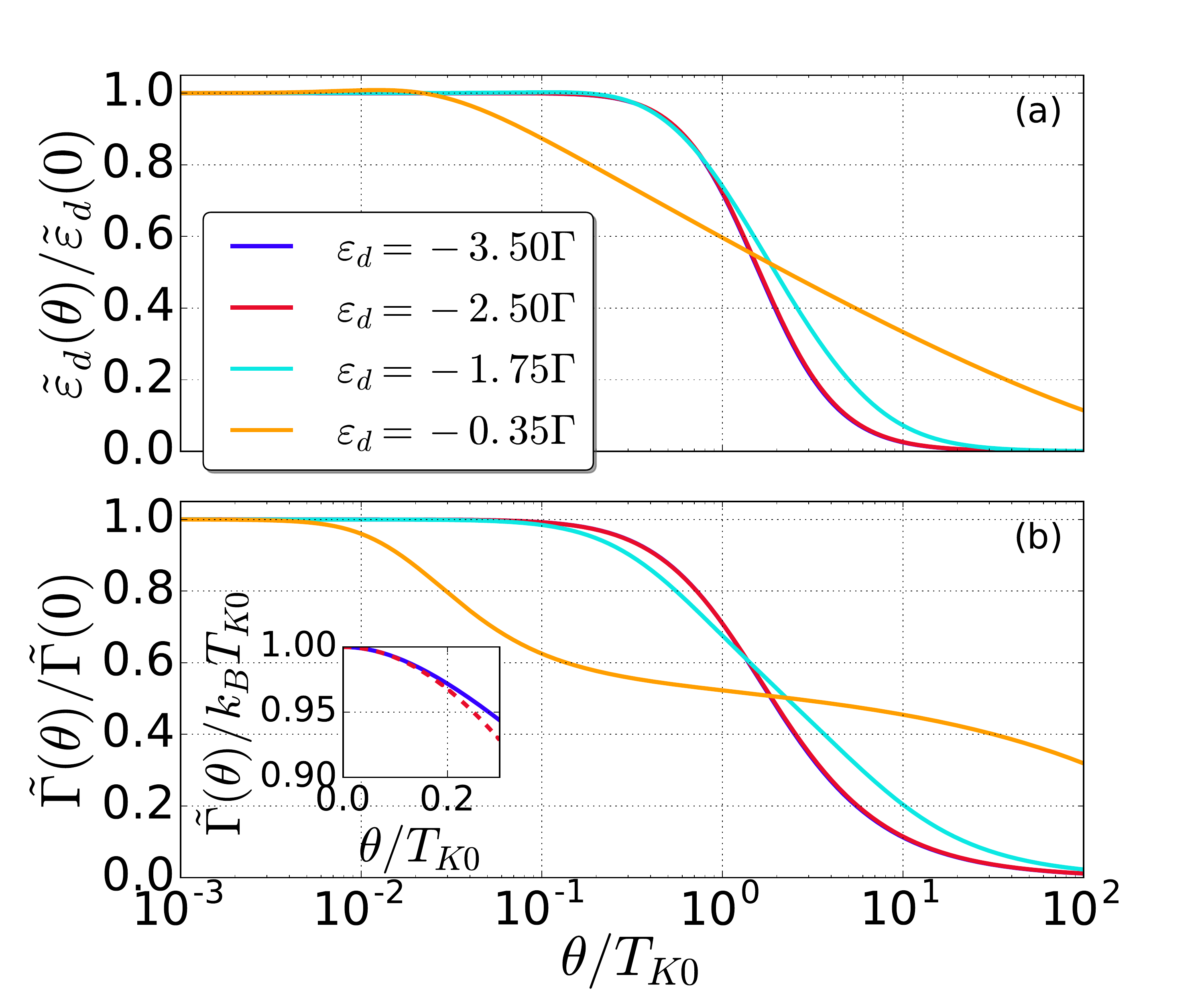}
\caption{(a) Renormalized dot gate position $\tilde{\varepsilon}_d$ and (b) resonance width $\tilde{\Gamma}$ as a function of the thermal bias $\theta$ for different $\varepsilon_d$ values within SBMFT. (Inset) Resonance width versus the thermal bias
from a numerical calculation (solid line) and from the analytical expression given by Eq. (\ref{TKvsT}) (dashed line) for  $\varepsilon_d=-3.5\Gamma$. Parameters: $D=100\Gamma$,  $k_B T= 0$, and $\Gamma_L=\Gamma_R=\Gamma/2$.}\label{fig:9}
\end{figure}

\begin{figure}[t!]
\centering
\includegraphics[width=0.5\textwidth,clip]{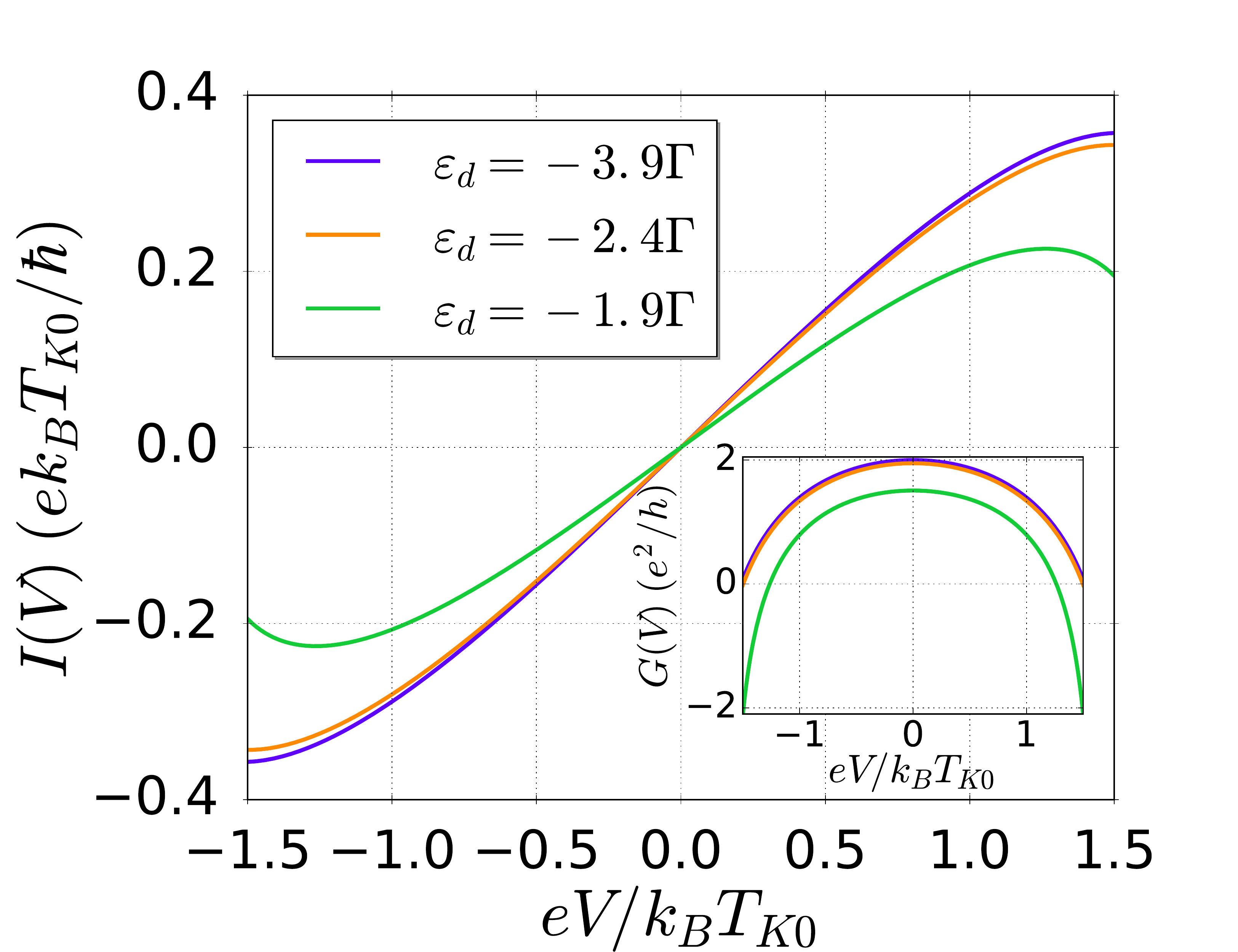}
\caption{Current-voltage characteristics of a single level quantum dot in the Kondo regime using slave-boson mean-field theory for different values of the gate voltage (level position). (Inset) Differential conductance of the quantum dot as a function of the applied voltage. Parameters:  $D=100\Gamma$,  $k_B T = 0$, and $\Gamma_L=\Gamma_R=\Gamma/2$}\label{fig:11}
\end{figure}

The solution of  the mean-field equations demonstrates that a thermal gradient, in contrast to a voltage bias, does not cause any Kondo splitting but merely renormalizes the Kondo resonance width. This renormalization is a decreasing function of the thermal bias $\theta$. We stress that this approximation is in principle valid only for low temperatures $T_L,T_R\ll T_{K0} $ for which SBMFT can be applied (see the inset of Fig.~\ref{fig:9}). Still, we arrive at the same conclusion as in the perturbation scheme of the previous section: a thermal bias gradually destroys the Kondo effect. The narrowing of the Abrikosov-Suhl resonance is the smoking gun of this quenching.

Our numerical results for $\theta>0$ are displayed in Fig.~\ref{fig:9}. Importantly, our analytical results agree well with the numerical curves provided we are in the deep Kondo regime ($\varepsilon_d=-3.5\Gamma$). Thus, as long as $\varepsilon_d$ enters the mixed-valence regime, we find departures from the analytical calculation, as expected. We observe three distinct regions for the $\tilde{\Gamma}(\theta)$ curves depending on the $\theta$ value. Thus, for 
a low temperature bias $\theta\ll 0.1T_{K0}$ the renormalized parameters remains almost unaffected. With increasing $\theta$ as $0.1T_{K0}<\theta<10T_{K0}$ the width of the Kondo peak decays exponentially and eventually drops to zero.

Surprisingly, the agreement between our perturbative analysis and the SBMFT result illustrated in Fig.~\ref{fig:13} is quite good. In both cases, we find a decrease of the binding energy of the many-body Kondo singlet as a function of the thermal bias. Notice that such agreement between both approaches holds even when the considered $\theta$ values are away from the range of validity for the SBMFT ($\theta$ small compared to $T_{K0}$) and the perturbative analysis ($\theta$ large compared to $T_{K0}$), which further supports the robustness of our main conclusion.

\subsection{Transport properties}\label{sec:LT-Transport}

Once the renormalized parameters are determined for a thermoelectric configuration the electrical current can be readily computed. We recall that the current is defined as the time derivative of the occupation of one of the leads $I_\alpha=-e\sum_{k\sigma}\partial_t \langle C_{\alpha k\sigma}^\dagger C_{\alpha k\sigma}\rangle$. Using the current conservation condition for steady-state currents ($I\equiv I_L=-I_R$) and considering the wide band limit, the current reads
\begin{equation}\label{EQI}
I=-\frac{e}{h}\int_{-\infty}^\infty d\omega \sum_\sigma \frac{4\Gamma_L\Gamma_R}{\Gamma}\text{Im}[G^r_{\sigma,\sigma}(\omega)][f_L(\omega)-f_R(\omega)]\, ,
\end{equation}
where $G^r_{\sigma,\sigma}(t,t')=-\frac{i}{\hbar}\theta(t-t')\langle [d_\sigma^\dagger(t'), d_\sigma(t)]_+\rangle$ is the QD retarded Green's function. Following the same procedure as above and performing the integration, one arrives at 
\begin{equation}\label{SlvCurrent}
I=I_0 \text{Im}\left[\psi\left(\frac{1}{2}+\frac{i(\tilde{\varepsilon}_d-\mu_R)+\tilde{\Gamma}}{2\pi k_BT_R}\right)-\psi\left(\frac{1}{2}+\frac{i(\tilde{\varepsilon}_d-\mu_L)+\tilde{\Gamma}}{2\pi k_BT_L}\right)\right]\; ,
\end{equation}
where $I_0=(8e\Gamma_L\Gamma_R)/(h\Gamma)$. In Fig.~\ref{fig:11}, the $I$-$V$ characteristic curves are shown for different dot gate positions.  At very low voltages the current is linear with the bias voltage as expected. From the $I$-$V$ curve it is easy to obtain the differential conductance which at zero bias can reach its maximum value (see the inset of Fig.~\ref{fig:11}).

\begin{figure}[b!]
\centering
\includegraphics[width=0.5\textwidth,clip]{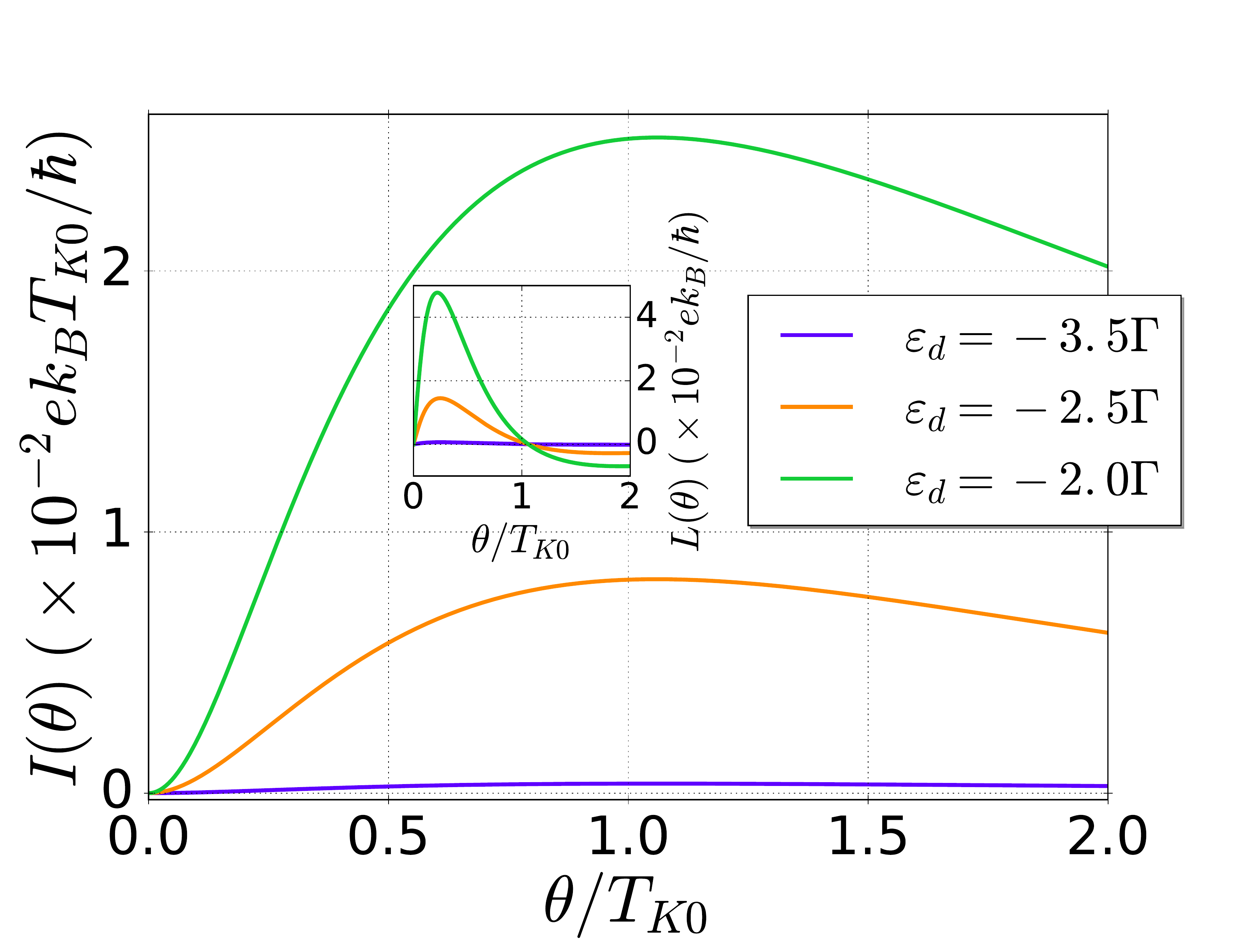}
\caption{Thermocurrent as a function the thermal gradient $\theta$ of a single level quantum dot in the Kondo regime using slave-boson mean-field theory for different values of the dot gate position. Inset: thermoelectric conductance as a function of the thermal bias for the same dot gate positions. Parameters: $D=100\Gamma$, $k_B T = 0$, and $\Gamma_L=\Gamma_R=\Gamma/2$}\label{fig:12}
\end{figure}

In Fig.~\ref{fig:12}  we show the results for the electrical current when a temperature shift is applied for different dot gate positions.  We observe that the thermocurrent attains higher values for dot level positions that are away from the deep Kondo regime, a regime in which electron-hole symmetry breaking is more prominent. If we express the current as $I\simeq L_1\theta + L_2 \theta^2$, particle-hole symmetry is responsible for vanishing thermoelectric conductance in linear response ($L_1\simeq 0$) since the transmission probability does not change significantly with the temperature bias. Then, the leading order at low thermal shifts is given by 
\begin{equation}\label{currentapprox}
L_2=\frac{4\pi^2 ek_B^2}{3h}
\tilde{\Gamma}_L\tilde{\Gamma}_R
\frac{\tilde{\varepsilon}_d}{\tilde{\varepsilon}_d^2+\tilde{\Gamma}^2}\; .
\end{equation}
Remarkably, the sign of $L_2$ in Eq.~(\ref{currentapprox}) depends on the renormalized dot level position. Thus we observe that deep in the Kondo regime ($\tilde{\varepsilon}_d\simeq 0$) the thermoelectric current is vanishingly small, even beyond linear response. This is better seen in the inset in Fig.~\ref{fig:12}, which depicts the nonlinear thermoelectric conductance $L=dI/d\theta$. The blue curve corresponds to the Kondo limit, which deviates little from the zero value. When the dot level shifts to higher energies closer to the Fermi energy the thermoelectric conductance behavior is more interesting. For instance, it shows a maximum value for small values of $\theta$ and then changes its sign. The maximal $L$
grows as the dot gate position enters the mixed-valence regime due to the lack of electron-hole symmetry.
The change of sign occurs at thermal biases of the order of the Kondo temperature because for larger values of $\theta$
the thermocurrent starts to decrease due to the quench of the Kondo effect (main panel of Fig.~\ref{fig:12}).

\section{Equation-of-motion technique}\label{Sec:NEGF}

%\subsection{Theoretical Model}\label{Sec:NEGFHamiltonian}

In this section, we address the thermoelectric transport in the Anderson model [Eq.~\eqref{eq_H}] for the case of moderate temperatures. In Sec.~III, we have employed the perturbative analysis, which is also suitable for high temperatures, to derive the Kondo temperature $T_K(\theta)$ for a finite temperature shift. However, perturbation theory does not allow us to access dynamical quantities like the local density of states. Therefore, in this section we propose an equation-of-motion (EOM) scheme that permits to study the role of single-particle peaks in addition to the Kondo resonance. In particular, we will be able to study the impact of the different resonances in the nonlinear transport regime. The EOM method aims at obtaining the retarded Green's function of the dot  $G^r_{\sigma,\sigma}$ within the Keldysh formalism \cite{mat610,Yme251,hewson2}. We find (the full calculation of $G^r_{\sigma,\sigma}$ is included in Appendix \ref{Ap:equationofmotion})
\begin{equation}\label{Gr}
\begin{split} G_{\sigma,\sigma}^r(\omega)=\frac{1-\langle \tilde{n}_{\bar{\sigma}} \rangle}{\omega-\varepsilon_d-\Sigma_0+U\Sigma_1/[\omega-\varepsilon_d-U-\Sigma_0-\Sigma_3]}\\
+\frac{\langle \tilde{n}_{\bar{\sigma}} \rangle}{\omega-\varepsilon_d-\Sigma_0-U-U\Sigma_2/[\omega-\varepsilon_d-\Sigma_0-\Sigma_3]} ,\end{split}
\end{equation}
where $\omega\to\omega + i 0^+$, $\bar{\sigma}$ denotes the spin opposite to $\sigma$ and
\begin{eqnarray}
\label{s0}\Sigma_0&=&\sum_{\alpha k}\frac{|\mathcal{V}_{\alpha k}|^2}{\omega-\varepsilon_{\alpha k }}, \\
\nonumber\Sigma_1&=&\sum_{\alpha k}\frac{\mathcal{V}_{\alpha k}^*}{\omega-\varepsilon_{\alpha k}}\left[\sum_{\beta q} (\mathcal{V}_{\beta q}\langle C_{\beta q \bar{\sigma}}^\dagger C_{\alpha k \bar{\sigma}}\rangle -\Sigma_0\langle d_{\bar{\sigma}}^\dagger C_{\alpha k \bar{\sigma}}\rangle)\right]\\
\nonumber &&+\sum_{\alpha k}\frac{\mathcal{V}_{\alpha k }}{-\omega_1+\varepsilon_{\alpha k}} \left[\sum_{\beta q} (\mathcal{V}_{\beta q}^*\langle C_{\alpha k \bar{\sigma}}^\dagger C_{\beta q \bar{\sigma}}\rangle \right.\\
\label{s1} &&\left.+\Sigma_0\langle C_{\alpha k \bar{\sigma}}^\dagger d_{\bar{\sigma}} \rangle)\right],\\
\label{s2}\Sigma_2&=&\Sigma_3-\Sigma_1 ,\\
\label{s3}\Sigma_3&=&\sum_{\alpha k }\left[\frac{|\mathcal{V}_{\alpha k}|^2}{\omega-\varepsilon_{\alpha k }}+\frac{|\mathcal{V}_{\alpha k }|^2}{\varepsilon_{\alpha k }-\omega_1}\right],\\
\nonumber\langle \tilde{n}_{\bar{\sigma}} \rangle&=&\langle n_{\bar{\sigma}} \rangle+\sum_{\alpha k}\frac{\mathcal{V}_{\alpha k}^*}{\omega-\varepsilon_{\alpha k}}\langle d_{\bar{\sigma}}^\dagger C_{\alpha k \bar{\sigma}}\rangle\\
\label{ntil}&&+\sum_{\alpha k}\frac{\mathcal{V}_{\alpha k}}{\omega_1-\varepsilon_{\alpha k}} \langle C_{\alpha k \bar{\sigma}}^\dagger d_{\bar{\sigma}} \rangle ,
\end{eqnarray}
with $\omega_1=-\omega+2\varepsilon_d+U$. Here, $\Sigma_0$ is the tunneling self-energy which in the wide-band limit is approximated as  $\Sigma_0\approx \Lambda(\omega)-i\Gamma$ where $\Lambda(\omega)$ is the principal value of $\Sigma_0$. $\Sigma_i$  ($i=1,2,3$) are additional self-energies which depend on the expectation values  $\langle d_{\bar{\sigma}}^\dagger C_{\alpha k \bar{\sigma}}\rangle$ and $\langle C_{\beta q \bar{\sigma}}^\dagger C_{\alpha k \bar{\sigma}}\rangle$. In Ref.~\cite{Yme251}, Meir \emph{et al.}~assume $\langle C_{\beta q \bar{\sigma}}^\dagger C_{\alpha k \bar{\sigma}}\rangle\approx f_\alpha(\omega)\delta_{\alpha\beta}\delta_{kq} \delta_{\sigma s}$ and $\langle d_{\bar{\sigma}}^\dagger C_{\alpha k \bar{\sigma}}\rangle\approx 0$. Nevertheless, in general the expectation values are functions of the lesser Green's function $G^<$ that should be calculated with the help of a modified fluctuation-dissipation theorem \cite{cab425} to account for the nonequilibrium situation 
\begin{equation}\label{flucdis}
\langle A^\dagger B \rangle =-\frac{1}{2\pi i}\int d\omega \mathcal{F}(\omega)(\llangle B, A^\dagger \rrangle^r -\llangle B, A^\dagger \rrangle^a) ,
\end{equation}
where the function $\mathcal{F}(\omega)$ is the effective non-equilibrium distribution function of the system [Eq.~(\ref{effetivefermi})] and $\llangle B, A^\dagger\rrangle^{r,a}$ is the Fourier transform of the retarded (advanced) correlator $\llangle B(t), A^\dagger(t')\rrangle^{r,a}=-\frac{i}{\hbar}\theta(t\mp t')\langle [A^\dagger (t') ,B(t)]_+\rangle$, $A$ and $B$ being two arbitrary second-quantization operators. Finally, Eq.~(\ref{ntil}) depends on the occupation of the quantum dot $\langle n_{\bar{\sigma}}\rangle$ and is calculated self-consistently from 
\begin{equation}\label{Ocupation}
\langle n_\sigma\rangle=\frac{1}{2\pi i}\int d\omega\; G^<_{\sigma,\sigma}(\omega)\; .
\end{equation}
Employing Eq.~\eqref{flucdis}, we write the lesser dot Green function as $G^<_{\sigma,\sigma}(\omega)=-2i\mathcal{F}(\omega)\text{Im}[G^r_{\sigma,\sigma}(\omega)]$.

\subsection{Infinite Coulomb interaction}

When $U$ is very large ($U\rightarrow \infty$), we immediately see that both $\Sigma_2$ and $\Sigma_3$ drop from Eq.~\eqref{Gr}. Therefore,  the dot Green's function reads
\begin{eqnarray}\label{Grinfty}
G_{\sigma,\sigma}^r(\omega)&=&\frac{1-\langle \tilde{n}_{\bar{\sigma}} \rangle}{\omega-\varepsilon_d-\Sigma_0-\Sigma_1}\,,
\end{eqnarray}
with %where $\langle \tilde{n}_{\bar{\sigma}} \rangle$ and $\Sigma_1$ are transformed to
\begin{align}
\label{s1infty}\Sigma_1&=\sum_{\alpha k}\frac{\mathcal{V}_{\alpha k}^*}{\omega-\varepsilon_{\alpha k}}\left[\sum_{\beta q} (\mathcal{V}_{\beta q}\langle C_{\beta q \bar{\sigma}}^\dagger C_{\alpha k \bar{\sigma}}\rangle -\Sigma_0\langle d_{\bar{\sigma}}^\dagger C_{\alpha k \bar{\sigma}}\rangle)\right],\\
\label{ntilinfty}\langle \tilde{n}_{\bar{\sigma}} \rangle&=\langle n_{\bar{\sigma}} \rangle+\sum_{\alpha k}\frac{\mathcal{V}_{\alpha k}^*}{\omega-\varepsilon_{\alpha k}}\langle d_{\bar{\sigma}}^\dagger C_{\alpha k \bar{\sigma}}\rangle\,,
\end{align}

The expectation values $\langle C_{\beta q \bar{\sigma}}^\dagger C_{\alpha k \bar{\sigma}}\rangle$ and $\langle d_{\bar{\sigma}}^\dagger C_{\alpha k \bar{\sigma}}\rangle$ in Eq.~\eqref{s1infty} and~\eqref{ntilinfty}
are evaluated using Eq.~(\ref{flucdis}). We next follow Ref.~\onlinecite{oen035}, which discusses a decoupling scheme for solving the set of EOM. By doing this,  Eqs.~(\ref{s1infty}) and (\ref{ntilinfty}) respectively become
\begin{eqnarray}
\Sigma_1(\omega)&=&-\frac{i\Gamma}{2}+\mathcal{X}(\omega)[1+2i\Gamma G^a_{\bar{\sigma},\bar{\sigma}}(\omega)]\Gamma,\\
\langle \tilde{n}_{\bar{\sigma}}\rangle &=&\langle n_{\bar{\sigma}}\rangle+\Gamma G^a_{\bar{\sigma},\bar{\sigma}}(\omega)\mathcal{X}(\omega),
\end{eqnarray}
where $\mathcal{X}(\omega)=\sum_\alpha \Gamma_\alpha X_\alpha(\omega)/\Gamma$ and $X_\alpha$ is defined as
\begin{eqnarray}
\nonumber X_\alpha(\omega)&=&\int_{-D}^D\frac{d\omega'}{\pi}\frac{f_\alpha(\omega')-1/2}{\omega-\omega'+i0^+}\\
\label{Xalpha}&=&\frac{1}{\pi}\left[\frac{1}{2}\ln{\frac{D^2-\omega^2}{(2\pi k_B T_\alpha)^2}}-\psi{\left(\frac{1}{2}-\frac{i(\omega-\mu_\alpha)}{2\pi k_BT_\alpha}\right)}\right]\,.
\end{eqnarray}
Due to the digamma function, Eq.~\eqref{Xalpha} contains logarithmic divergences that are responsible for the emergence of the Kondo singularity. Finally, the QD retarded Green's function takes the form
\begin{eqnarray}
\label{GrUinftyfinal}G_{\sigma,\sigma}^r(\omega)=g(\omega)\left[\delta n_\sigma+\frac{iQ(\omega)}{\mathcal{X}^*(\omega)}\right],
\end{eqnarray}
with 
\begin{eqnarray}
\label{gomega}g(\omega)&=&\frac{1}{\omega-\varepsilon_d-\Lambda +i3\Gamma/2},\\
Q(\omega)&=&S(\omega)-\sqrt{S^2(\omega)+|\mathcal{X}(\omega)|^2\left(\frac{3}{2} \delta n_\sigma-\delta n_\sigma^2\right)},\\
\label{Somega}S(\omega)&=&z^2+\frac{9}{16}-z\text{Re}{[\mathcal{X}(\omega)]+\left(\delta n_\sigma-\frac{3}{4}\right)\text{Im}{[\mathcal{X}(\omega)]}},
\end{eqnarray}
where $\delta n_\sigma=1-\langle n_\sigma \rangle$
and $z=(\omega-\varepsilon_d-\Lambda)/2\Gamma$. Equation (\ref{GrUinftyfinal}) has two different terms.
The first term in the right-hand side depends only on $\delta n_\sigma$ and is responsible for the dot mean-field resonance. The second term is responsible for the Kondo singularity that becomes prominent for $T<T_{K0}$.  

Once we have determined the retarded dot Green's function, we can analyze the behavior of the local density of states. First, we estimate the
intrinsic Kondo temperature from Haldane's formula \cite{fdm416} 
\begin{equation}\label{KondoT}
T_{K0}=\sqrt{D\Gamma}e^{-\pi|\varepsilon_d|/2\Gamma} \; .
\end{equation}
As expected, a Kondo peak is visible in the spectral function $\rho_d(\omega)=\sum_\sigma\rho_{d\sigma}(\omega)$ defined as 
\begin{equation}\label{SpectralEOM}
\rho_{d\sigma}=-\frac{1}{\pi}\text{Im}{[G^r_{\sigma,\sigma}(\omega)]}\; .
\end{equation} 
This narrow resonance will survive as long as $T<T_{K0}$ and, by increasing $T$, smears out until it completely disappears \cite{oen035}. 

\begin{figure}[t!]
\centering
\includegraphics[width=0.50\textwidth,clip]{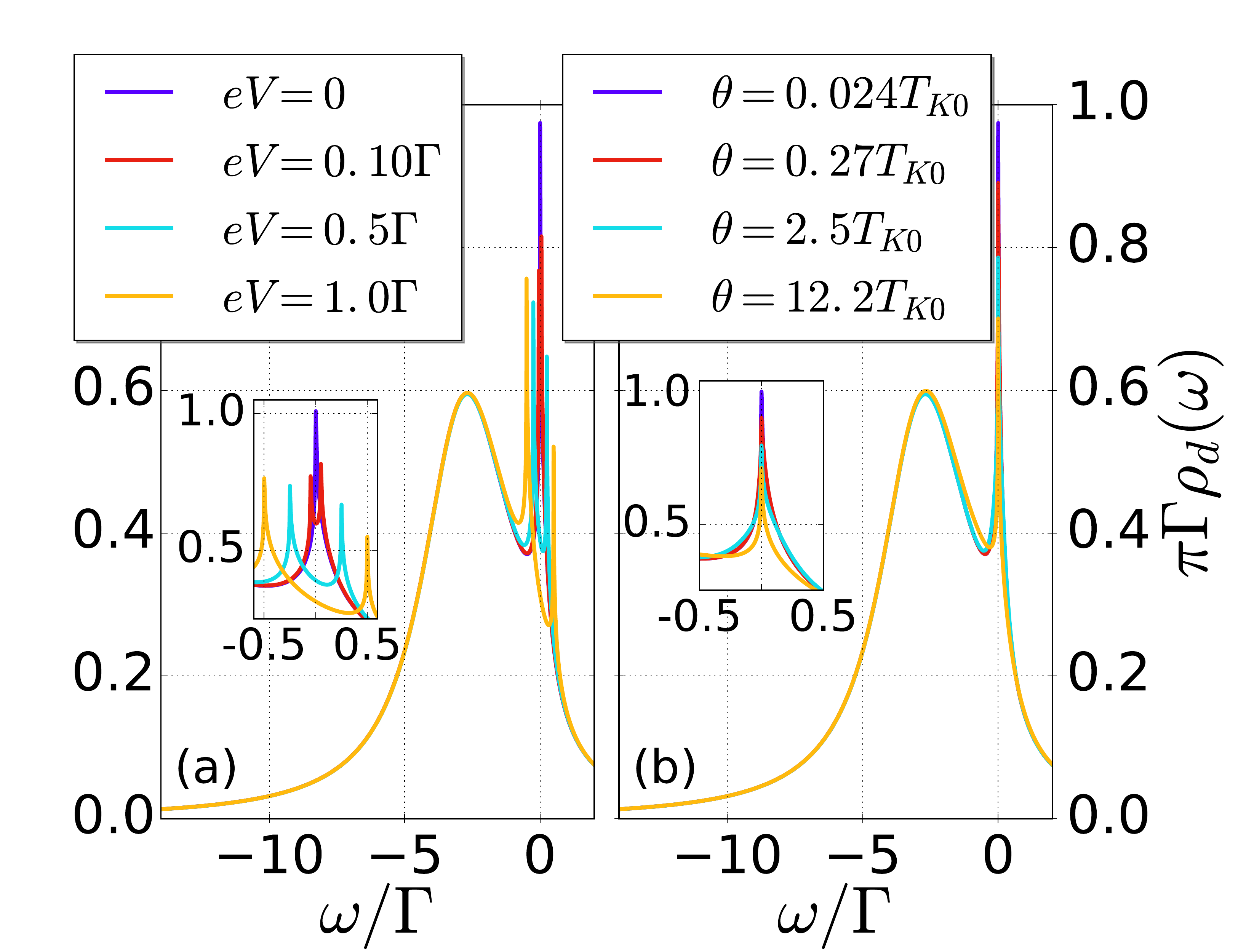}
\caption{(a) Nonequilibrium infinite-$U$ quantum dot spectral density of states for different $eV$ values. Inset:  Detail of the density of states around the Fermi energy ($\varepsilon_F=0$).  (b) Nonequilibrium infinite-$U$ quantum dot spectral density of states for different thermal gradients. The background temperature is set at $T=0.024 T_{K0}$. Inset:  Detail of the density of states around the Fermi energy ($\varepsilon_F=0$). Parameters: $\varepsilon_d=-3.5\Gamma$,  $D=100\Gamma$,  $T=0.024 T_{K0}$.}\label{fig:2}
\end{figure}

The impact of voltage and temperature biases on the spectral function is illustrated in Fig. \ref{fig:2}. First, for the voltage-driven case
%with $\mu_L=-\mu_R=eV/2$,
the Kondo resonance becomes split with resonances $\omega\approx\pm eV/2$ [Fig.\ \ref{fig:2}(a)],
as experimentally observed~\cite{silvano}. Yet, the employed EOM scheme does not capture the dephasing effect generated by the voltage shift. A combination of EOM using the dot occupation computed with the non-crossing approximation has been  proposed to amend the lack of dephasing when a voltage is considered \cite{aro156,yme260,cab425}.  For the temperature-driven case, we recall our thermal configuration in which only the left contact is heated, $T_L=T+\theta$, whereas the right reservoir is kept at the background temperature, $T_R=T$. Our results are depicted in Fig.~\ref{fig:2}(b).  As expected, the main effect of the thermal gradient is to smear out the Kondo singularity. However, a further increase of the temperature difference does not cause the Kondo peak to vanish even for $\theta\gg T_{K0}$ [see Fig.~\ref{fig:2}(b)]. This can be understood as follows. The Kondo resonance in EOM arises from the sharp character of the Fermi function. In this case, even though one of the Fermi functions becomes sufficiently smooth as $\theta$ increases, the other contact Fermi function remains sharp for a low background temperature. As a consequence, even if $\theta$ grows considerably the Kondo peak is never totally quenched. This behavior is an artifact of the EOM approach since it does not account for dephasing processes as we earlier pointed out. However, this method gives the correct behavior at low $\theta$: a thermal bias does not split the Kondo resonance but gradually destroys the peak, in agreement with the perturbative approach and the mean-field slave-boson theory.
 
\subsection{Finite Coulomb interaction}

In this section, we generalize our previous findings to the case where the Coulomb interaction is finite.
In this case, Eqs.~(\ref{s1}), (\ref{s3}) and (\ref{ntil}) take the following forms
\begin{eqnarray}
\nonumber \Sigma_1&=&-i\Gamma+\Gamma\mathcal{X}(\omega)[1+2i\Gamma G^a_{\bar{\sigma},\bar{\sigma}}(\omega)]\\
&&-\Gamma\mathcal{X}^*(\omega_1)[1+2i\Gamma G^r_{\bar{\sigma},\bar{\sigma}}(\omega_1)],\\
\Sigma_3&=&\Lambda(\omega)+\Lambda(\omega_1)-2i\Gamma,\\
\nonumber \langle \tilde{n}_{\bar{\sigma}} \rangle&=&\langle n_{\bar{\sigma}} \rangle+\Gamma[G^a_{\bar{\sigma},\bar{\sigma}}(\omega)\mathcal{X}(\omega)\\
&&-G^r_{\bar{\sigma},\bar{\sigma}}(\omega_1)\mathcal{X}(\omega_1)]\; .
\end{eqnarray}
Note that now the retarded Green's function depends recursively on itself, which makes the decoupling scheme a highly nontrivial task. In order to make further progress we consider a finite but large $U$ interaction. Hence, we can safely neglect $G^r(\omega_1)$ and find an explicit expression for the following retarded Green's function
\begin{eqnarray}\label{GrUlargefinal}
G_{\sigma,\sigma}^r(\omega)=g_u(\omega)\left[\delta n_u+\frac{iQ_u(\omega)}{\mathcal{X}_u^*(\omega)}\right],
\end{eqnarray}
The quantities entering in Eq.~(\ref{GrUlargefinal}) depend on the function  $u(\omega)=U/(\Sigma_0+\Sigma_3+\varepsilon_d+U-\omega)$:
\begin{eqnarray}
\mathcal{X}_u(\omega)&=&u(\omega)\mathcal{X}(\omega),\\
g_u(\omega)&=&\frac{1}{2\Gamma(z+i(1+u)/2+\bar{\mathcal{X}}_u)},\\
\nonumber Q_u(\omega)&=&S_u-\left({S_u^2-|\mathcal{X}^*_u\delta n_u|^2}\right.\\
\label{Eq:Qu}&&\left.{+|\mathcal{X}_u|^2\left[\delta n_uh_1(\omega)-2\text{Im}{[\delta n_u]}h_2(\omega)\right]}\right)^{1/2},\\
\nonumber S_u(\omega)&=&z^2+\frac{|1+u|^2}{4}-\frac{\text{Im}{[\bar{\mathcal{X}}_u(1+u)]}}{2}+\frac{|\bar{\mathcal{X}}_u|^2}{4}\\
\nonumber &&-z\text{Re}{[\mathcal{X}_u]}-\frac{\text{Im}{[\mathcal{X}_u(1+u^*)]}}{2}-\frac{\text{Re}{[\mathcal{X}_u\bar{\mathcal{X}}_u]}}{2}\\
&& +\text{Im}{[\mathcal{X}_u \delta n_u^*]}-z\text{Im}{[u]}+z\text{Re}{[\bar{\mathcal{X}}_u]},
\end{eqnarray}
where the bar indicates that the function is to be computed as  $\bar{f}\equiv f(\omega_1)$. In Eq.~(\ref{Eq:Qu}) we have defined the functions $h_1(\omega)=-\text{Im}{[\bar{\mathcal{X}_u}]}+1+\text{Re}{[u]}$,  $h_2(\omega)=z+i(1+u)/2+\bar{\mathcal{X}}_u^*/2$ and $\delta n_u=1-u(\omega)\langle n_\sigma \rangle$.   

\begin{figure}[t!]
\centering
\includegraphics[width=0.5\textwidth,clip]{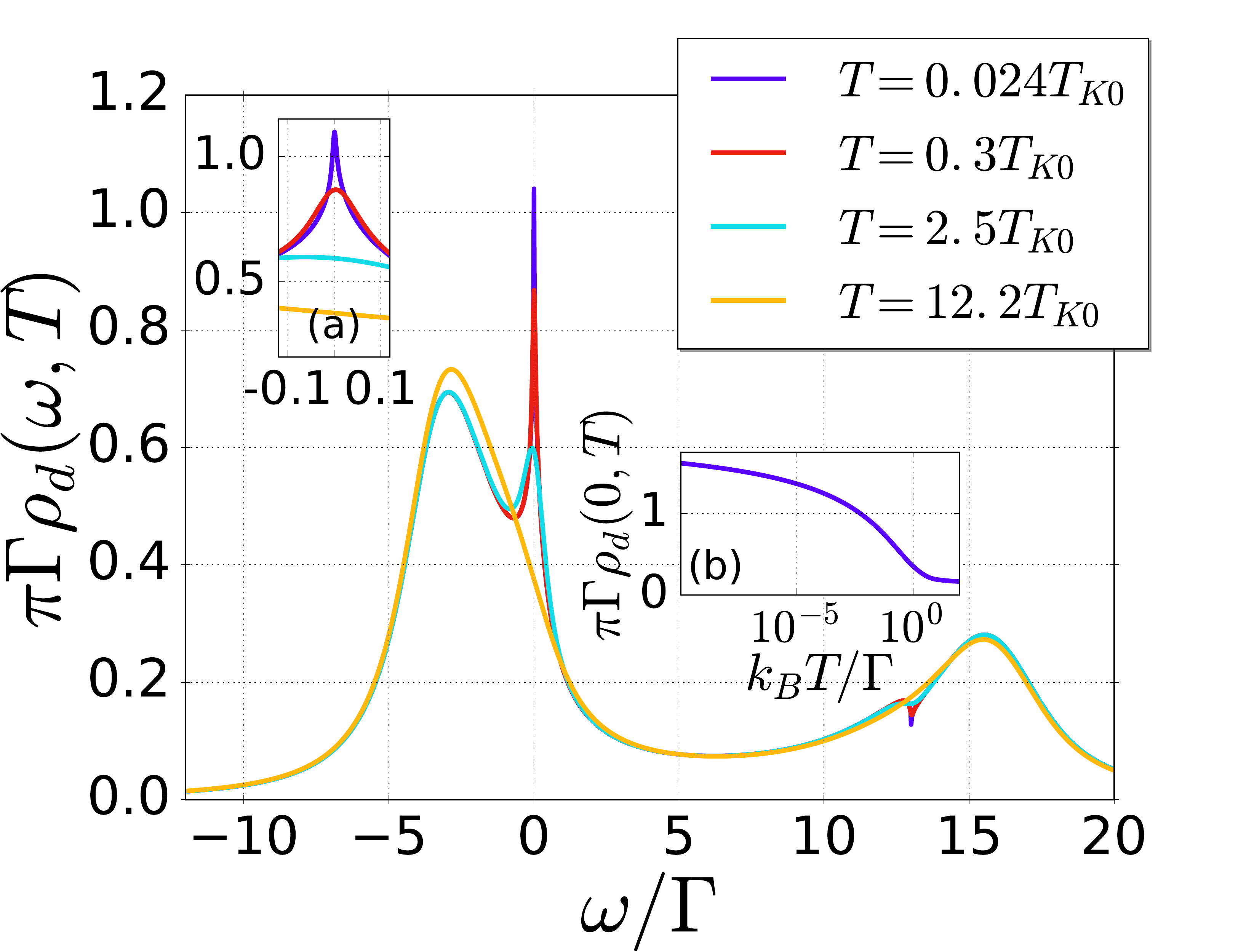}
\caption{Finite-$U$ quantum dot spectral density at equilibrium for different background temperatures. Parameters: $\varepsilon_d=-3.5\Gamma$,  $D=100\Gamma$, $U=20\Gamma$. Insets: (a) Detail of the dot spectral density of states around the Fermi energy. (b) Height of the Kondo peak as a function of the background temperature. }\label{fig:finiteU}
\end{figure}

Figure \ref{fig:finiteU} shows our results for the dot spectral density [Eq.~(\ref{SpectralEOM})] at high background temperatures. As expected, the dot DOS consists of two mean-field resonances, one centered at $\omega= \varepsilon_d$ and another at $\omega=\varepsilon_d+U$ and the Kondo singularity at $\omega \approx 0$.  For these results we take into account a modified Kondo temperature for a system with finite $U$ \cite{hewson}
\begin{equation}\label{KondoTU}
T_{K0}\approx\sqrt{2\Gamma U}\exp{\left\{-\frac{\pi|\varepsilon_d|(U+\varepsilon_d)}{2\Gamma U}\right\}}.
\end{equation}
We observe in Fig.~\ref{fig:finiteU} that the Kondo peak is quenched as long as the background temperature surpasses the Kondo temperature, as expected. Additionally, the inset of Fig.~\ref{fig:finiteU} shows the amplitude of the Kondo peak as a function of $T$. This result confirms the fact that a high temperature destroys the Kondo singularity as long as $T>T_{K0}$ even when we consider finite charging energies. We have also extended our study of the dot DOS for finite $U$ to the nonequilibrum case by considering the influence of a voltage and temperature gradient. We observe qualitatively similar results (not shown here) as those obtained for the infinite $U$ case [Fig.~\ref{fig:2}]. We find that given voltage leads to a Kondo peak splitting whereas a temperature gradient smears out the Kondo singularity. 

\subsection{Voltage-driven transport} 

\begin{figure}[t]
\centering
\includegraphics[width=0.5\textwidth,clip]{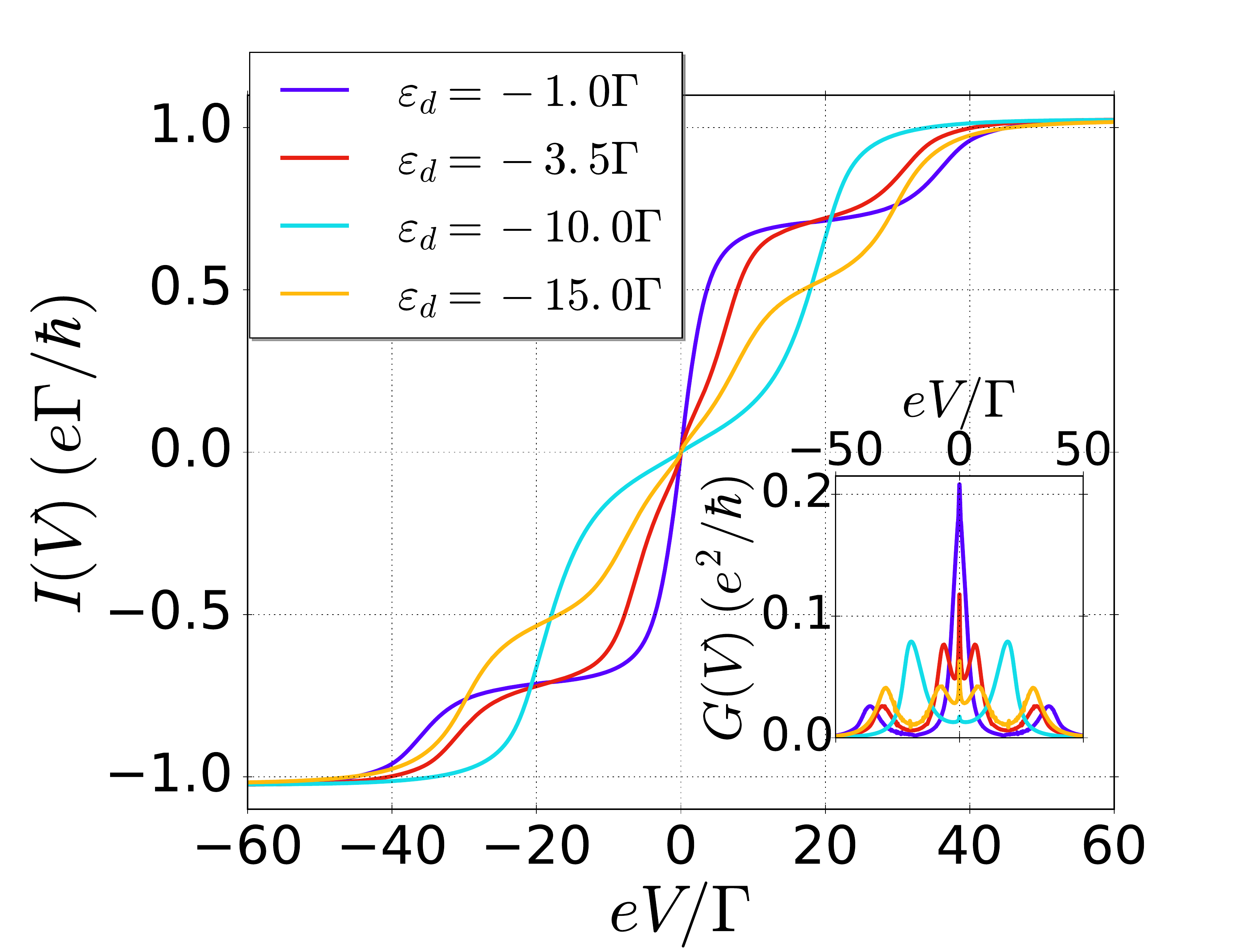}
\caption{$I$-$V$ characteristics at different dot level positions. (Inset) Differential electric condutance vs voltage bias for the given values of the energy level. Parameters:  $D=100\Gamma$,  $k_B T=0.0001\Gamma$,  $U=20\Gamma$, and $\Gamma_L=\Gamma_R=\Gamma/2$.}\label{fig:5}
\end{figure}

We proceed to the transport properties of the system. We calculate the current from Eq.~(\ref{EQI}) using Eq.~(\ref{GrUlargefinal}). The integrals over energy of the self-consistent calculation are now solved numerically. Figure \ref{fig:5} shows the $I$--$V$ characteristic for different dot gate positions. The data display a staircase-like behavior where the step transitions take place whenever the bias voltage aligns with the dot resonances. By direct differentiation, we find the differential conductance curves shown in the inset of Fig.~\ref{fig:5}. Here, when $\varepsilon_d<-\Gamma$ we find five different peaks, four located when the mean field resonances aligns with the electrochemical potential of the leads [$eV\approx\pm 2\varepsilon_d$ and $eV\approx\pm 2(\varepsilon_d+U)$] and the Kondo zero bias anomaly centered at equilibrium ($V=0$).  Whenever a resonance occurs the $I$-$V$ has a visible jump. Our calculation is performed at finite temperature. Therefore, the Kondo peak is apparent in the differential conductance only for sufficiently negative dot gate positions. 

\subsection{Temperature-driven transport}

\begin{figure}[t!]
\centering
\includegraphics[width=0.5\textwidth,clip]{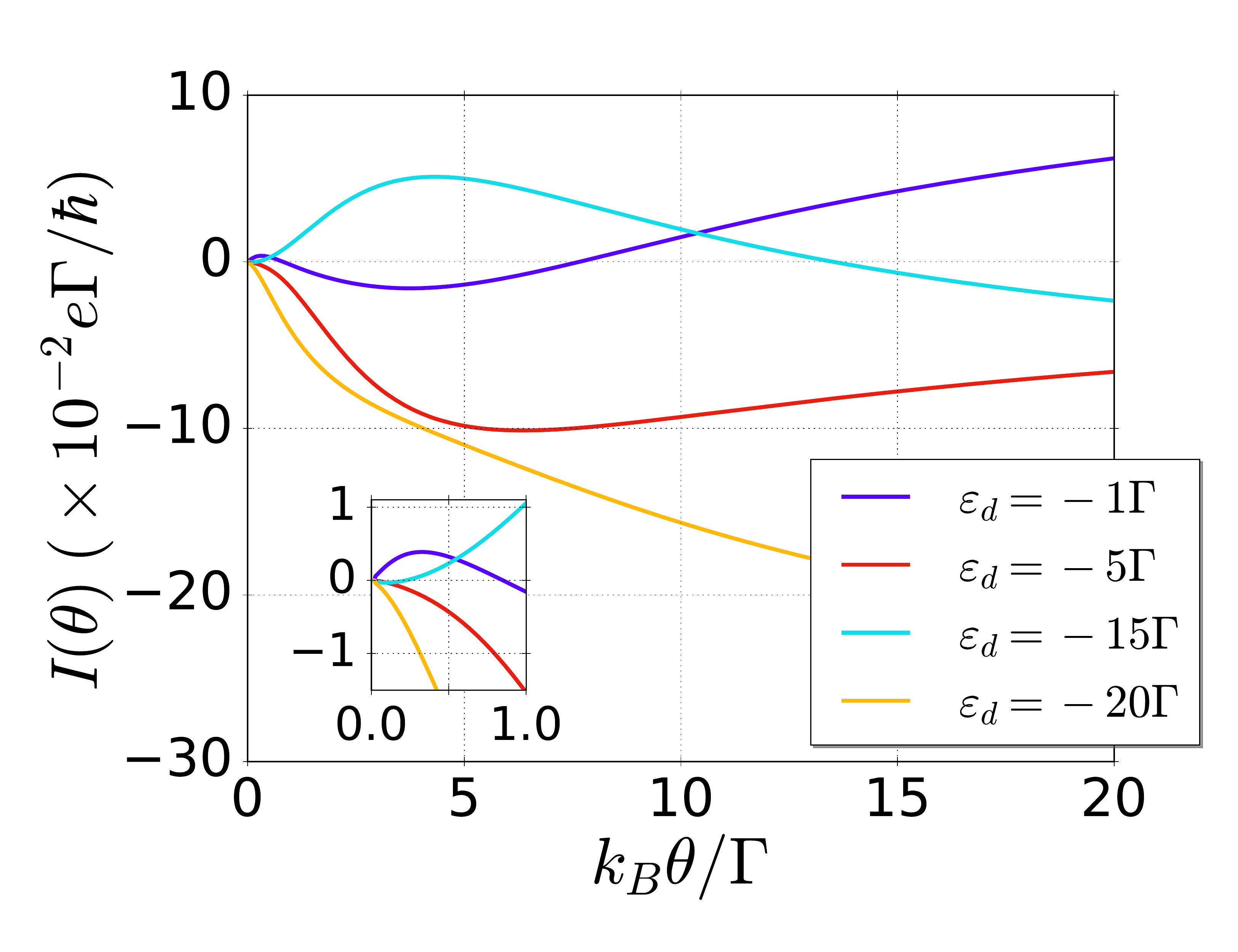}
\caption{Thermocurrent vs the thermal gradient for different dot gate positions.  (Inset) Detail of the thermocurrent at low thermal gradients. Parameters: $U=20\Gamma$, $D=100\Gamma$,  $k_B T=0.01\Gamma$, and $\Gamma_L=\Gamma_R=\Gamma/2$}\label{fig:7}
\end{figure}
We consider now the thermoelectric response of the electrical current in Fig.\ \ref{fig:7}.
The thermocurrent $I(\theta)$ is calculated for the thermal configuration sketched in Fig.~\ref{fig:0}. 
Depending on the energy level $\varepsilon_d$ we find different regions where the thermocurrent behaves distinctively: if $\varepsilon_d>0$ or $\varepsilon_d + U <0$ we find that the dot resonances are either above (empty orbital regime) or below (full orbital regime) the Fermi energy. Then, for these dot gate positions the thermocurrent either decreases or increases monotonously (e.g., in Fig.~\ref{fig:7} for $\varepsilon_d=-20\Gamma$). Otherwise, when $0>\varepsilon_d>-U$ the $I_{\rm th}(\theta)$ curves change sign. Notably, the nontrivial zeros occur at different energy scales determined by either spin fluctuations ($k_BT_{K0}$), see inset of Fig.\ \ref{fig:7}, or charge fluctuations ($U$). See, e.g., the case $\varepsilon_d=-\Gamma$ with a nontrivial zero at around $k_B \theta=\Gamma$ and another zero at around $k_B \theta=10\Gamma$. Notice that for very negative dot gate positions the nontrivial zero of the thermoelectric current associated with the Kondo scale may not occur due to the fact that for those level values the Kondo temperature is exceedingly small and one has $T>T_{K0}$. 
\begin{figure*}[t!]
\centering
\includegraphics[width=1\textwidth,clip]{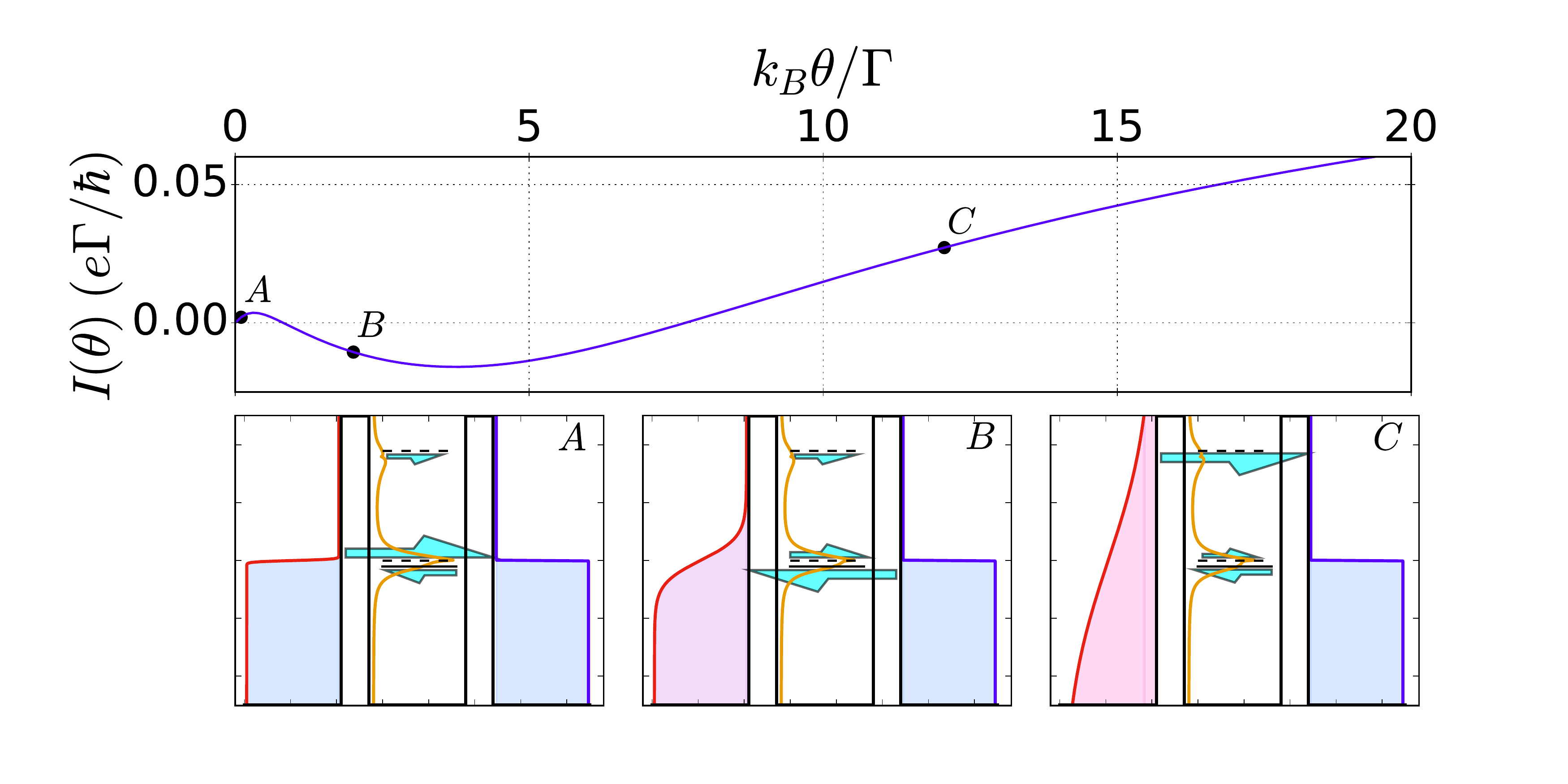}
\caption{(Top) Thermocurrent vs $\theta$ for $\varepsilon_d=-\Gamma$ as taken from Fig.~\ref{fig:7}. Bottom: Energy diagram corresponding to the current states marked in the upper panel. Red (blue) line indicates the Fermi-Dirac function of the left (right) reservoir where orange curve corresponds to the spectral function for the three points ($A$, $B$ and $C$) indicated in the top panel. $\varepsilon_d$ is indicated with a black line. Dashed lines corresponds to the $\varepsilon_F=0$ and $\varepsilon_d+U$ energies. Finally, the arrows show the direction and intensity of the electronic flow through the corresponding resonance. }\label{fig:7b}
\end{figure*}
The nontrivial zeros can be explained with the aid of an energy diagram. In Fig.\ \ref{fig:7b} (top panel)
we reproduce $I(\theta)$ for $\varepsilon_d=-\Gamma$ and plot the dot configurations in the bottom panel for three points indicated above.
The QD shows three resonances located approximately at  $\omega\approx\varepsilon_d$, $\omega\approx	\varepsilon_d+U$ (single-particle peaks), and the Kondo peak at $\omega\simeq\varepsilon_F=0$. In the very low-temperature regime, the only open transport channel is due to the Kondo resonance through which electrons flow between both leads. Away from the particle-hole symmetry point, the Kondo peak becomes asymmetric and lies slightly above $\varepsilon_F$. Then, if we heat the left reservoir up, electrons tend to flow from the left to the right side  (case $A$ of Fig.\ \ref{fig:7b}). Increasing $\theta$, the single-particle resonance at $\varepsilon_d$ starts to contribute to the current flow but in the opposite direction and eventually it dominates over the Kondo resonance. In between (case $B$), for a given value of $\theta$ the current vanishes. Further increasing of $\theta$ opens the higher single-particle resonance at $\varepsilon_d+U$, which favors electron tunneling from left to right, i.e., opposite to the previous current flow. As a consequence, an additional sign change of the thermoelectrical current takes place (case $C$). 

\section{Conclusions}\label{Sec:Conclusions}

In closing, we have examined the nonequilibrium thermoelectric effects of a correlated quantum dot connected to two electronic reservoirs. We have employed different theoretical approximations with different ranges of validity.  As a first attempt, we have applied the perturbative analysis to the Kondo Hamiltonian when a thermal and a voltage bias is present. Interestingly, we find that the Kondo temperature decreases monotonically in the presence of a thermal gradient when one reservoir is heated. At very low temperatures, we have employed the infinite-$U$ slave-boson mean-field theory suitable for the Fermi liquid regime. Here, we obtain a very good agreement with the perturbative results for the behavior of $T_K$ when a thermal shift is considered.  Finally, in order to investigate the density of states in the high and moderate temperature regime we have applied the equation-of-motion scheme to our setup. By using this approach we are able to  capture both Coulomb blockade and Kondo physics. We consider both infinite and finite charging energies.
We observe that the Kondo peak splits under the action of a voltage bias. In the case of a thermal gradient, the Kondo peak decreases slowly.
Finally, we have analyzed both the $I$--$V$ characteristic and the thermoelectric current. For the latter we find the existence of nontrivial zeros at two distinct energy scales ($k_B\theta$), each associated with Kondo correlations and charge fluctuations.   We explain the existence of these nontrivial zeros due to a change of the flow caused by the contribution of the different resonances as long as $\theta$ grows.

Admittedly, our employed methods have a limited range of validity. For instance, none of them properly captures dephasing, which should
play a significant role for finite voltage and temperature biases, as pointed out in Sec.\ V. 
In the presence of dephasing, the strongly correlated Kondo state lacks full coherence and the peak conductance then decreases.
One possibility to fix this issue is to attach a fictitious voltage probe that mimics dephasing processes in the dot~\cite{three}.
We expect a stronger quenching of the Kondo resonance. However, further work is needed to investigate dephasing effects
from more microscopic models.

Our work contributes to the understanding of the fate of the Kondo effect in thermally driven quantum dots far from equilibrium.
We have provided key theoretical predictions that might be experimentally tested with present transport techniques. 

\section{Acknowledgments}
We thank R. Aguado for useful discussions. This work has been supported by MINECO under Grants FIS2014-52564, CAIB and FEDER. 

%\newpage 
\appendix

\section{Calculation of the conductance}\label{Ap:ScalingPert}
The goal of this appendix is to find an expression for the electrical conductance using the expectation value of Eq.~(\ref{Eq:ISmatrix}). We start our calculation performing a perturbation expansion around the coupling constant $\mathcal{J}_{\alpha \beta}^{(0)}$ (see Ref.~\cite{kam815}). To first order in perturbation theory, the electrical current is absent $\langle \hat{I}(0)\rangle =0$. Therefore, we need to start at second order where the current, after some algebraic manipulations, takes the form
\begin{align}\label{Eq:I2order}
\nonumber I^{(2)}= \left(-\frac{2}{\hbar}\right) \sum_{\vec{\alpha} \vec{\beta}} \mathcal{J}_{\alpha \beta}^{(0)}&\text{Im}{ \bigg[ \int_{-\infty}^0 dt e^{-\frac{ie}{\hbar}(V_\alpha-V_\beta)t} }\\
&\times \langle \bar{T}[x_{\sigma s} C_{\vec{\alpha}}^\dagger C_{\vec{\beta}} \hat{I}(0)] \rangle\bigg]\; ,
\end{align}
where $\bar{T}$ is the anti-time-ordering operator.
To make the notation more compact, we have defined in Eq.~\eqref{Eq:I2order}
the sets $\vec{\alpha}\equiv\{\alpha,k,\sigma\}$ and $\vec{\beta}\equiv\{\beta,q,s\}$.
Inserting Eq.~(\ref{Eq:Ioperator}) into Eq.~(\ref{Eq:I2order}) and applying Wick's theorem to the expectation values,
the electrical current turns out to be
\begin{align}\label{Eq:I2order2}
\nonumber I^{(2)}&= -2e   \left[\mathcal{J}_{LR}^{(0)}\right]^2 \left(\sum_{\sigma_1 \sigma_2} \langle \bar{T}[x_{\sigma_1 \sigma_2} x_{\sigma_2 \sigma_1}]\rangle\right)\sum_\alpha(1-2\delta_{\alpha R})\\
\times &\int_{-\infty}^0 dt \text{Re}{\left[\sum_{k_1k_2} e^{-\frac{ie}{\hbar}(V_\alpha-V_{\bar{\alpha}})t}g^{\bar{t}}_{\alpha k_1}(0,t)g_{\bar{\alpha} k_2}^{\bar{t}}(0,t)\right]}\; ,
\end{align}
where $g^{\bar{t}}_{\alpha k_1}(t_1,t_2)$ is the anti-time-ordered Green's function of the free electrons in lead $\alpha$. It reads
\begin{equation}\label{Eq:atGreen}
g^{\bar{t}}_{\alpha k_1}(t_1,t_2)=-\frac{i}{\hbar}e^{-\frac{i}{\hbar}\varepsilon_{\alpha k_1} (t_1-t_2)}\left[\theta(t_2-t_1)-f_\alpha(\varepsilon_{\alpha k_1})\right]\; .
\end{equation}
Using the identity $\sum_{\sigma_1\sigma_2}\langle \bar{T}[x_{\sigma_1 \sigma_2} x_{\sigma_2 \sigma_1}]\rangle=1/2$, substituting Eq.~(\ref{Eq:atGreen}) into Eq.~(\ref{Eq:I2order2}) and solving the integral in time and in $k_i$, we find
\begin{eqnarray}\label{Eq:I2nO}
I^{(2)}= -\frac{e^2\pi}{\hbar}\nu^2 \left[\mathcal{J}_{LR}^{(0)}\right]^2 V\;.
\end{eqnarray}
Therefore, the second-order conductance term becomes constant independently of temperature. Hence we need to compute the third order in perturbation theory:
\begin{widetext}
\begin{align}
\nonumber I^{(3)}&=  -\frac{1}{\hbar^2} \sum_{\vec{\alpha}_i \vec{\beta}_i} \mathcal{J}_{\alpha_1 \beta_1}^{(0)} \mathcal{J}_{\alpha_2 \beta_2}^{(0)}\text{Re}{\left[\int_{-\infty}^0 dt_2\int_{-\infty}^0 dt_1 e^{-\frac{ie}{\hbar} (V_{\alpha_1}-V_{\beta_1}) t_1}e^{-\frac{ie}{\hbar} (V_{\alpha_2}-V_{\beta_2}) t_2}\langle \bar{T}[x_{\sigma_1 s_1} C_{\vec{\alpha}_1}^\dagger C_{\vec{\beta}_1}x_{\sigma_2 s_2}C_{\vec{\alpha}_2}^\dagger C_{\vec{\beta}_2}\hat{I}(0)]\rangle\right]}\\
\label{Eq:A5}&+\frac{1}{\hbar^2} \sum_{\vec{\alpha}_i \vec{\beta}_i} \mathcal{J}_{\alpha_1 \beta_1}^{(0)} \mathcal{J}_{\alpha_2 \beta_2}^{(0)}\text{Re}{\left[\int_{-\infty}^0 dt_2\int_{-\infty}^0 dt_1 e^{-\frac{ie}{\hbar} (V_{\alpha_1}-V_{\beta_1}) t_1}e^{-\frac{ie}{\hbar} (V_{\alpha_2}-V_{\beta_2}) t_2}\langle x_{\sigma_1 s_1} C_{\vec{\alpha}_1}^\dagger C_{\vec{\beta}_1}\hat{I}(0)x_{\sigma_2 s_2}C_{\vec{\alpha}_2}^\dagger C_{\vec{\beta}_2}\rangle\right]}\;.
\end{align}

Now, we separate both terms in the right hand side of Eq.~(\ref{Eq:A5}) as $I^{(3a)}$ and $I^{(3b)}$. We proceed in the same way as the second order and apply Wick's theorem. After lenghty but straightforward algebra, we find

\begin{eqnarray}
\nonumber I^{(3a)}&=&e \sum_{\alpha\beta } \left[\mathcal{J}_{LR}^{(0)}\right]^2 \mathcal{J}_{\alpha \alpha}^{(0)}\left(\sum_{\sigma_i} \langle \bar{T}[x_{\sigma_1 \sigma_2}(t_1) x_{\sigma_2 \sigma_3} (t_2) x_{\sigma_3 \sigma_1}(0)]\rangle+\langle \bar{T}[x_{\sigma_1 \sigma_2}(t_2) x_{\sigma_3 \sigma_1} (t_1) x_{\sigma_2 \sigma_3}(0)]\rangle \right)(1-2\delta_{\beta L})\\
 &\times&\int_{-\infty}^0 dt_1 \int_{-\infty}^0 dt_2\text{Re}{\left[\sum_{k_i}e^{-\frac{ie}{\hbar} (V_\beta-V_\alpha)t_1}e^{-\frac{ie}{\hbar} (V_\alpha-V_{\bar{\beta}})t_2} g^{\bar{t}}_{\beta k_1}(0,t_1)g^{\bar{t}}_{\alpha k_2}(t_1,t_2)g^{\bar{t}}_{\bar{\beta} k_3}(t_2,0)\right]}\;,\label{eqA6}\\
\nonumber I^{(3b)}&=&-e \sum_{\alpha \beta } \left[\mathcal{J}_{LR}^{(0)}\right]^2 \mathcal{J}_{\alpha \alpha}^{(0)} \left(\sum_{\sigma_i} \langle x_{\sigma_1 \sigma_2} x_{\sigma_2 \sigma_3} x_{\sigma_3\sigma_1}\rangle \right) \int_{-\infty}^0 dt_1 \int_{-\infty}^0 dt_2(1-2\delta_{\beta L}) \text{Re}{\left[ \sum_{\{k_i\}} e^{-\frac{ie}{\hbar}(V_\alpha - V_{\bar{\beta}})t_1} e^{-\frac{ie}{\hbar}(V_{\beta}-V_\alpha)t_2}\right.}\\
\nonumber &\times& {\left.g_{\beta k_1}^>(0,t_2) g^<_{\alpha k_2}(t_2,t_1) g^>_{\bar{\beta} k_3}(t_1,0) \right]} -e \sum_{\alpha \beta } \left[\mathcal{J}_{LR}^{(0)}\right]^2 \mathcal{J}_{\alpha \alpha}^{(0)} \left(\sum_{\sigma_i} \langle x_{\sigma_1 \sigma_2} x_{\sigma_3 \sigma_1} x_{\sigma_2\sigma_3}\rangle \right) \\
&\times & \int_{-\infty}^0 dt_1 \int_{-\infty}^0 dt_2(1-2\delta_{\beta R}) \text{Re}{\left[ \sum_{k_i} e^{-\frac{ie}{\hbar}(V_\beta - V_{\alpha})t_1} e^{-\frac{ie}{\hbar}(V_{\alpha}-V_{\bar{\beta}})t_2} g_{\beta k_1}^<(0,t_1) g^>_{\alpha k_2}(t_1,t_2) g^<_{\bar{\beta} k_3}(t_2,0) \right]}\;,\label{eqA7}
\end{eqnarray}

where $g^<_{\alpha k}$ and $g^>_{\alpha k }$ are the lesser and greater Green's function for electrons on lead $\alpha$. The spin expectation values of Eq.~\eqref{eqA6} depends on the time ordering and yield $5/8+(3/8)\text{sgn}(t_2-t_1)$, with $\text{sgn}(t)$ the sign function. Meanwhile, the terms in Eq.~\eqref{eqA7} are independent  of time and the spin expectation values are thus $1/8$ and $1/2$, respectively. Substituting these values and the definition of the Green's function in Eq.~(\ref{Eq:atGreen}),
\begin{eqnarray}
g^<_{\alpha k \sigma}(t_1,t_2)&=&\frac{i}{\hbar}e^{-\frac{i}{\hbar}\varepsilon_{\alpha k}(t_1-t_2)}f_\alpha(\varepsilon_{\alpha k }) \; ,\\
g^>_{\alpha k \sigma}(t_1,t_2)&=&-\frac{i}{\hbar}e^{-\frac{i}{\hbar}\varepsilon_{\alpha k}(t_1-t_2)}(1-f_\alpha(\varepsilon_{\alpha k})) \; ,
\end{eqnarray}
the current up to third order reads
\begin{eqnarray}
\nonumber I^{(3)}&=&\frac{e}{8\hbar^3} \sum_{\alpha\beta } \left[\mathcal{J}_{LR}^{(0)}\right]^2 \mathcal{J}_{\alpha \alpha}^{(0)} (1-2\delta_{\beta L})\int_{-\infty}^0  dt_1\int_{-\infty}^0 dt_2 \left\{(5-3\text{sgn}(t_2-t_1))\right. \\
\nonumber &\times & \text{Im}{\left[\sum_{k_i} e^{-\frac{i}{\hbar} [\varepsilon_{\alpha k_2}-\varepsilon_{\beta k_1} +e(V_{\beta}-V_{\alpha})]t_1} e^{-\frac{i}{\hbar} [\varepsilon_{\bar{\beta} k_3}-\varepsilon_{\alpha k_2} +e(V_{\alpha}-V_{\bar{\beta}})]t_2}f_{\beta}(\varepsilon_{\beta k_1})(\theta(t_2-t_1)-f_\alpha (\varepsilon_{\alpha k_2}))(1-f_{\bar{\beta}}(\varepsilon_{\bar{\beta} k_3})) \right]}\\
\nonumber &-& \text{Im}{\left[\sum_{k_i}e^{-\frac{i}{\hbar}[\varepsilon_{\alpha k_2}-\varepsilon_{\beta k_1}+e(V_\beta-V_\alpha)]t_1}e^{-\frac{i}{\hbar}[\varepsilon_{\bar{\beta}k_3}-\varepsilon_{\alpha k_2} + e(V_\alpha-V_{\bar{\beta}})]t_2} \left( f_{\alpha}(\varepsilon_{\alpha k_2})-f_\beta(\varepsilon_{\beta k_1})f_{\alpha}(\varepsilon_{\alpha k_2})-f_{\bar{\beta}}(\varepsilon_{\bar{\beta}k_3}) f_{\alpha}(\varepsilon_{\alpha k_2})\right.\right.}\\
\label{Eq:A8}&-&\left.\left.\left.4f_{\bar{\beta}}(\varepsilon_{\bar{\beta}k_3}) f_{\beta}(\varepsilon_{\beta k_1})+5f_{\bar{\beta}}(\varepsilon_{\bar{\beta}k_3}) f_{\alpha}(\varepsilon_{\alpha k_2})f_{\beta}(\varepsilon_{\beta k_1}) \right)\right]\right\}.
\end{eqnarray}
We next combine the Fermi functions of Eq.~(\ref{Eq:A8}) and perform the sums over the wavenumbers $k_i$ by transforming them into integrals. We solve the resulting integrals by performing the Fourier transform of the Fermi function:
\begin{eqnarray}
\int_{-\infty}^{\infty} d\omega \frac{e^{-i\omega t}}{1+e^{\hbar \omega/k_B T}}=\frac{\pi i}{\sinh{\frac{\pi k_B t T}{\hbar}}}\; .
\end{eqnarray}
Then, the current becomes
\begin{eqnarray}
\nonumber I^{(3)}&=&-\frac{e\pi^3}{8\hbar^2}\nu^3 \left[\mathcal{J}_{LR}^{(0)}\right]^2 \sum_{\alpha\beta}\mathcal{J}^{(0)}_{\alpha \alpha} (1-2\delta_{\beta L})\int_{-\infty}^0 dt \left[3\text{Bs}_{\beta\bar{\beta}}(t)-\text{Bs}_{\beta\alpha}(t)+\text{Bs}_{\alpha\bar{\beta}}(t)\right]\\
&&-\frac{e\pi^3}{8\hbar^2}\nu^3 \left[\mathcal{J}_{LR}^{(0)}\right]^2 \sum_{\alpha\beta}\mathcal{J}^{(0)}_{\alpha \alpha} (1-2\delta_{\beta L})\text{Re}{\left[\int_{-\infty}^0 dt_2 e^{-\frac{ie}{\hbar}(V_\alpha-V_{\bar{\beta}})t_2}\int_{-\infty}^0 dt_1 e^{-\frac{ie}{\hbar}(V_\beta-V_{\alpha})t_1}  \text{Ts}_{\beta\alpha\bar{\beta}}(t_1,t_2)\right]}\;,
\end{eqnarray}
where we define the functions
\begin{eqnarray}
\text{Bs}_{\alpha\gamma}(t)&= \frac{\sin{[e(V_\alpha-V_\gamma)t/\hbar]}}{\beta_\alpha \beta_\gamma \sinh{\frac{\pi t}{\beta_\alpha \hbar}}\sinh{\frac{\pi t}{\beta_\gamma \hbar}}}\;,
\end{eqnarray}
\begin{align}
\text{Ts}_{\alpha\gamma\delta}(t_1,t_2)&= \frac{\text{sgn}(t_2-t_1)}{\beta_\alpha \beta_\gamma \beta_\delta\sinh{\frac{\pi t_1}{\beta_\alpha \hbar}} \sinh{\frac{\pi (t_1-t_2)}{\beta_\gamma \hbar}} \sinh{\frac{\pi t_2}{\beta_\delta \hbar}}}\; ,
\end{align}
with $\beta_\alpha=1/k_B T_\alpha$ the inverse temperature of lead $\alpha$.
Finally, after solving the sum over $\alpha$ and $\beta$ we obtain
\begin{eqnarray}\label{Eq:I3rO}
\nonumber I^{(3)}&=&\frac{3e\pi^3}{4\hbar^2}\nu^3\left[\mathcal{J}_{LR}^{(0)}\right]^2 \left(\mathcal{J}^{(0)}_{LL}+\mathcal{J}^{(0)}_{RR}\right)\\
&\times&\int_{-\infty}^0 dt \frac{\sin{(eVt/\hbar)}}{\beta_L \beta_R \sinh{\frac{\pi t}{\beta_L\hbar}}\sinh{\frac{\pi t}{\beta_R \hbar}}}\;.
\end{eqnarray}
Once we have the third order term, we are interested in the linear conductance $G=\partial I/\partial V |_{V=0}$, which defines the height of the Kondo resonance. Applying the voltage derivative to Eq.~(\ref{Eq:I2nO}) and Eq.~(\ref{Eq:I3rO}) the conductance is given by
\begin{eqnarray}
\nonumber G&=& -\frac{e^2\pi}{\hbar}\nu^2 \left[\mathcal{J}^{(0)}_{LR}\right]^2+\frac{3e^2\pi^3}{4\hbar^3}\nu^3\left[\mathcal{J}_{LR}^{(0)}\right]^2 \left(\mathcal{J}^{(0)}_{LL}+\mathcal{J}^{(0)}_{RR}\right) \\
&\times& \int_{-\infty}^0 dt \frac{t}{\beta_L \beta_R \sinh{\frac{\pi t}{\beta_L\hbar}}\sinh{\frac{\pi t}{\beta_R \hbar}}+D_0^{-2}}\; .
\label{eq_Gintegral}
\end{eqnarray}
We note that we have added the energy bandwidth $D_0=\sqrt{-\varepsilon_d(U+\varepsilon_d)}$ to the integral in order to find a convergent solution. Additionally, we need to assume that the minimum time of the integral in Eq.~\eqref{eq_Gintegral} is related to the temperatures as $\tau\approx \hbar \sqrt{\beta_L \beta_R}$. These assumptions yield Eq.~\eqref{conductance}.

\section{Equation of motion beyond Hartree-Fock}\label{Ap:equationofmotion}
The aim of this appendix is to give more details of the equation-of-motion calculation yielding Eq. (\ref{Gr}). Firstly, we compute the equation of motion for the retarded Green's function 
\begin{equation}\label{Cor1}
(i\hbar \partial_t-\varepsilon_d) G^r_{\sigma,\sigma}(t,t')=\delta(t-t')+\sum_{\alpha k} \mathcal{V}_{\alpha k} G^r_{\alpha k \sigma,\sigma} + U\llangle d_\sigma n_{\bar{\sigma}} , d_\sigma^\dagger \rrangle\, .
\end{equation}
Equation (\ref{Cor1}) depends on $\llangle d_\sigma n_{\bar{\sigma}} , d_\sigma^\dagger \rrangle$, which is a higher-order correlator. Considering that $U\gg k_B T, \Gamma$ we calculate the equation of motion for $\llangle d_\sigma n_{\bar{\sigma}} , d_\sigma^\dagger \rrangle$,
\begin{eqnarray}
\label{Cor2}(i\hbar \partial_t -\varepsilon_d-U)\llangle d_\sigma n_{\bar{\sigma}} , d_\sigma^\dagger \rrangle=\langle n_{\bar{\sigma}} \rangle \delta(t-t')+\sum_{\alpha k} \mathcal{V}_{\alpha k}^*\left[\llangle C_{\alpha k\sigma} n_{\bar{\sigma}} , d_\sigma^\dagger \rrangle+\llangle d_\sigma d_{\bar{\sigma}}^\dagger C_{\alpha k \bar{\sigma}} , d_\sigma^\dagger \rrangle\right]-\sum_{\alpha k} \mathcal{V}_{\alpha k}\llangle d_\sigma  C_{\alpha k \bar{\sigma}}^\dagger d_{\bar{\sigma}} , d_\sigma^\dagger \rrangle .
\end{eqnarray}
At this point, if we determine the EOM of $\llangle C_{\alpha k\sigma} n_{\bar{\sigma}} , d_\sigma^\dagger \rrangle$ and  neglect the contributions of the correlators $\llangle d_\sigma d_{\bar{\sigma}}^\dagger C_{\alpha k \bar{\sigma}} , d_\sigma^\dagger \rrangle$, $\llangle d_\sigma  C_{\alpha k \bar{\sigma}}^\dagger d_{\bar{\sigma}} , d_\sigma^\dagger \rrangle$, $\llangle C_{\alpha k \sigma}  C_{\beta q \bar{\sigma}}^\dagger d_{\bar{\sigma}} , d_\sigma^\dagger \rrangle$ and $\llangle C_{\alpha k\sigma} d_{\bar{\sigma}}^\dagger C_{\beta q \bar{\sigma}} , d_\sigma^\dagger \rrangle$ we arrive at the two peak solution that  has been widely investigated \cite{hew144}. However, our interest for the moment is to take a step further and include Kondo correlations. We follow the calculation of
Lacroix \cite{lac238} and Kashcheyevs \emph{et al.} \cite{kas125} and extend the EOM scheme by computing the evolution of these higher-order correlators:
\begin{eqnarray}
\label{Cor3}(i\hbar \partial_t -\varepsilon_{\alpha k})\llangle C_{\alpha k\sigma} n_{\bar{\sigma}} , d_\sigma^\dagger \rrangle&=&\mathcal{V}_{\alpha k}\llangle d_{\sigma} n_{\bar{\sigma}} , d_\sigma^\dagger \rrangle+\sum_{\beta q} \left[\mathcal{V}_{\beta q}^* \llangle C_{\alpha k\sigma} d_{\bar{\sigma}}^\dagger C_{\beta q \bar{\sigma}} , d_\sigma^\dagger \rrangle-\mathcal{V}_{\beta q} \llangle C_{\alpha k\sigma} C_{\beta q \bar{\sigma}}^\dagger d_{\bar{\sigma}}, d_\sigma^\dagger \rrangle\right] , \\
\nonumber (i\hbar \partial_t -\varepsilon_{\alpha k})\llangle d_\sigma d_{\bar{\sigma}}^\dagger C_{\alpha k \bar{\sigma}} , d_\sigma^\dagger \rrangle&=&\langle d_{\bar{\sigma}}^\dagger C_{\alpha k \bar{\sigma}} \rangle \delta(t-t')+\mathcal{V}_{\alpha k}\llangle d_{\sigma} n_{\bar{\sigma}} , d_\sigma^\dagger \rrangle\\
\label{Cor4}&& +\sum_{\beta q} \left[\mathcal{V}_{\beta q}^* \llangle C_{\beta q\sigma} d_{\bar{\sigma}}^\dagger C_{\alpha k \bar{\sigma}} , d_\sigma^\dagger \rrangle-\mathcal{V}_{\beta q}\llangle d_\sigma C_{\beta q\bar{\sigma}}^\dagger C_{\alpha k \bar{\sigma}}  , d_\sigma^\dagger \rrangle\right],\\
\nonumber (i\hbar \partial_t +\delta \varepsilon_{\alpha k})\llangle d_\sigma  C_{\alpha k \bar{\sigma}}^\dagger d_{\bar{\sigma}} , d_\sigma^\dagger \rrangle&=&  \langle  C_{\alpha k \bar{\sigma}}^\dagger d_{\bar{\sigma}}\rangle\delta(t-t')-\mathcal{V}_{\alpha k}^*\llangle d_{\sigma} n_{\bar{\sigma}}  , d_\sigma^\dagger \rrangle+\\
\label{Cor5} && +\sum_{\beta q} \mathcal{V}_{\beta q}\left[\llangle C_{\beta q\sigma}  C_{\alpha k \bar{\sigma}}^\dagger d_{\bar{\sigma}} , d_\sigma^\dagger \rrangle+\llangle d_\sigma  C_{\alpha k \bar{\sigma}}^\dagger C_{\beta q\bar{\sigma}} , d_\sigma^\dagger \rrangle\right] ,
\end{eqnarray}

where $\delta \varepsilon_{\alpha k}=\varepsilon_{\alpha k}-2\varepsilon_d-U$. New higher-order correlators are generated in the process. Then, in order to obtain a solvable system of differential equations, we consider the approximation proposed by Mattis \cite{ath978,kas125}:
\begin{equation}\label{Matisapprox}
\llangle A^\dagger BC, D^\dagger \rrangle\approx \langle A^\dagger B \rangle \llangle C , D^\dagger \rrangle-\langle A^\dagger C \rangle \llangle B, D^\dagger\rrangle \;.
\end{equation}
Therefore, after applying Eq.~(\ref{Matisapprox}) to Eqs.~(\ref{Cor3}), (\ref{Cor4}) and (\ref{Cor5}), the system of equations can be closed. In the Fourier space, we end up with Eq.~(\ref{Gr}).
\end{widetext}

\end{document}